\documentclass[10pt,twocolumn,letterpaper]{article}

\usepackage{iccv}
\usepackage{times}
\usepackage{epsfig}
\usepackage{graphicx}
\usepackage{amsmath}
\usepackage{amssymb}
\usepackage[english]{babel}

\usepackage{booktabs,caption,siunitx}
\usepackage[dvipsnames]{xcolor} 
\usepackage{wrapfig}
\usepackage{subcaption}
\usepackage{floatrow}
\usepackage{caption}
\usepackage{microtype}

\newfloatcommand{capbtabbox}{table}[][\FBwidth]

\usepackage[pagebackref=true,breaklinks=true,colorlinks,bookmarks=false]{hyperref}

\makeatletter
\@namedef{ver@everyshi.sty}{}
\makeatother
\usepackage{tikz}
\usepackage{pgfplots}
\usepackage{graphicx}
\usepackage{subcaption}
\usepackage{capt-of}
\usepackage{comment}
\usepackage{color}
\usepackage{booktabs}
\usepgfplotslibrary{fillbetween}
\pgfplotsset{compat=1.17}
\usepackage{duckuments} 
\definecolor{c_b0}{HTML}{A4DEF9}
\definecolor{c_b0.1}{HTML}{D2CCA1}
\definecolor{c_b0.5}{HTML}{387780}
\definecolor{c_b1}{HTML}{E83151}
\definecolor{c_b5}{HTML}{CFBAE1}
\definecolor{c_rnn}{HTML}{313715}
\definecolor{c_vpbs}{HTML}{E83151}
\definecolor{c_vgbs}{HTML}{0096C7}
\definecolor{c_tracktor}{HTML}{000000}
\definecolor{c_centertrack}{HTML}{000000}
\definecolor{c_kalman}{RGB}{0,50,255}

\iccvfinalcopy 

\def\iccvPaperID{4183} 

\ificcvfinal\fi

\newcommand{\rita}[1]{{\color{Blue}{Rita: #1}}}

\begin{document}

\title{PixInWav: Residual Steganography for Hiding Pixels in Audio}

\author{Margarita Geleta$^1$ \:\: Cristina Punt\'{i}$^{1,2}$ \:\: Kevin McGuinness$^2$\\
Jordi Pons$^3$ \:\:Cristian Canton$^1$ \:\: Xavier Giro-i-Nieto$^{1,4,5}$
\and
\small$^1$\emph{Universitat Polit\`{e}cnica de Catalunya} \quad $^2$\emph{Dublin City University} \quad 
\small$^3$\emph{Dolby Labs}\\
\small$^4$\emph{Institut de Robòtica i Informàtica Industrial, CSIC-UPC} \quad
\small$^5$\emph{Barcelona Supercomputing Center} 
}

\maketitle
\ificcvfinal\thispagestyle{empty}\fi

\begin{abstract}
Steganography comprises the mechanics of hiding data in a host media that may be publicly available. While previous works focused on unimodal setups (e.g., hiding images in images, or hiding audio in audio), PixInWav targets the multimodal case of hiding images in audio. To this end, we propose a novel residual architecture operating on top of short-time discrete cosine transform (STDCT) audio spectrograms. Among our results, we find that the residual audio steganography setup we propose allows independent encoding of the hidden image from the host audio without compromising quality. Accordingly, while previous works require both host and hidden signals to hide a signal, PixInWav can encode images offline --- which can be later hidden, in a residual fashion, into any audio signal. Finally, we test our scheme in a lab setting to transmit images over airwaves from a loudspeaker to a microphone verifying our theoretical insights and obtaining promising results.

\end{abstract}

\section{Introduction}

\par Steganography (from Greek, “steganós” meaning covered, and “graphein” meaning writing) refers to the method of concealing a \textit{container} signal embedding a \textit{hidden} signal within a \textit{host} signal. The resulting {container} signal may be sent through a publicly accessible channel in a way that the {hidden} signal stays inconspicuous to potential observers. 
Steganography has benefited from recent advances in deep learning, especially in the uni-modal front: image/video~\cite{tancik2020stegastamp, zhu2018hidden} or audio~\cite{kreuk2020hide}. In this work, we focus on the unexplored multi-modal case of hiding images in audio signals, such that the {host} signal is audio and the {hidden} signal is an image.

\begin{figure}[!t]
\begin{center}
   \includegraphics[width=1\linewidth]{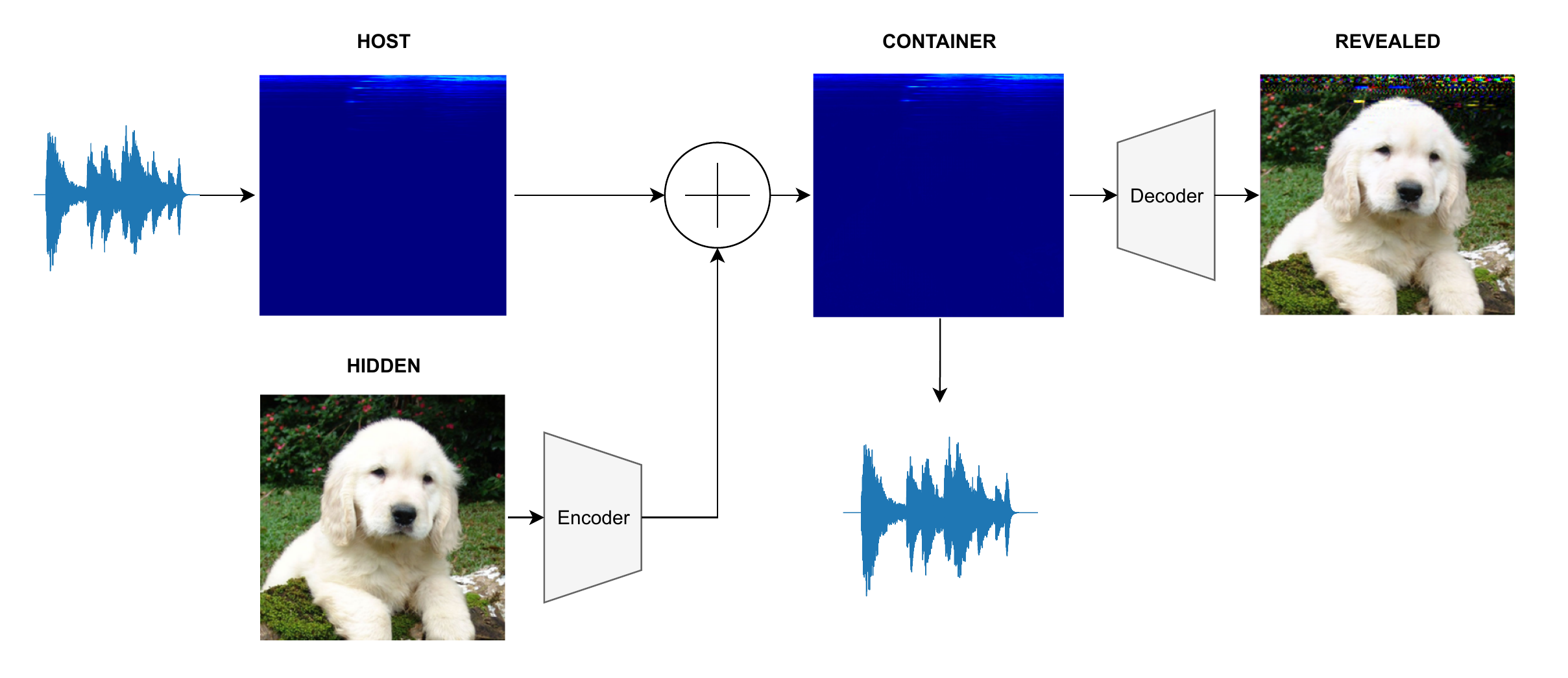}
\end{center}
   \caption{PixInWav proposes learning independent image representations that can be hidden within audio signals.}
\label{fig:teaser}
\end{figure}

\par Hiding images into audio allows the exploitation of existing audio infrastructures for image and video distribution. For instance, analog broadcast radio may transport the album cover of a song being played, loudspeakers in airports distribute maps or visual messages for the hearing impaired, or video streams distributed to the handheld devices of crowds located in areas with insufficient mobile network capacity.
The proposed technical solution also has direct applications to watermarking, where the hidden information is related to its content thus enabling potential provenance solutions, and media forensics, where digital content must be analyzed to determine whether it is authentic, fake, or if it has been modified, so that backdoor attacks can be prevented~\cite{li2020invisible}. 

Deep neural networks have attracted the attention of steganography researchers, as they have the potential to learn the best representations for hiding the {secret} signal within the {host}. 
Unlike many popular steganographic classic methods that encode the hidden signal within the least significant bits of the host signal, deep learning approaches can compress and spread the secret signal’s representation over all the available bits.
Previous deep learning approaches~\cite{baluja2019hiding,duan2019reversible,tancik2020stegastamp,zhu2018hidden} have adopted neural architectures in which both the {host} and the hidden signals are fed into a deep neural network to construct the container signal.
Our solution does not rely on this costly merge operation. Instead, our proposed \textit{PixInWav} model relies on a simple (yet effective) residual architecture~\cite{he2016deep}: we train a neural encoder for the hidden image independently from the {host}, so that a fixed and learned representation can be simply added to the audio spectrogram to build the container signal~---~see Figure \ref{fig:teaser}. Throughout our work, we show that it is not necessary to learn a different representation for each hidden signal depending on the host, since the same encoded representation of the secret can be added to any audio.

Our contributions can be summarized as follows:
(a)~we explore, for the first time, sending images over audio waveforms; 
(b)~we address this task via a novel residual architecture operating on top of the short-time discrete cosine transform (STDCT) audio representation; 
and (c)~we show that encoded images be created independently from the audio signal, such that they can be pre-computed per image and later added to any arbitrary audio signal.

\section{Related Work}


Deep steganography has only been applied in unimodal setups (e.g., images or audio, separately).
In the following, we review these two research directions to position our multimodal work, {PixInWav}, within the state-of-the-art.

\vspace*{-3mm}
\paragraph{Deep Image Steganography.}

Hiding information in image pixels has been extensively addressed in the broad field of steganography, with significant progress achieved by recent deep learning techniques~\cite{UED,WOW, S-UNIWARD, pevny2010using, DBLP:journals/corr/abs-1907-06956,LSB}.
Most deep image steganography techniques rely on convolutional neural network (CNN) encoders to hide the message within a host image, and a CNN decoder to recover the hidden message.
Such systems are generally trained following a loss schema in which (i) the encoder is trained to minimize a reconstruction loss over the host image; and (ii) the decoder is trained to minimize the reconstruction loss over the (recovered) secret image.
As a result, the output of the encoder (the container) includes an image that is perceptually similar to the host image but contains a (hidden) secret image --- that the decoder can recover.




\textit{Hidden}~\cite{zhu2018hidden} encodes the message by replicating and concatenating it with the pixel embeddings obtained by a 2D convolutional encoder. {PixInWav}, instead, simply adds the message directly into the host image in a residual fashion. \textit{Hidden} also incorporates an adversarial loss to improve the realistic appearance of the container. 
\textit{Stegastamp}~\cite{tancik2020stegastamp} follows a similar approach, but focuses on the specific application of encoding hyperlinks in image pixels, reporting satisfactory results for up to 56 bits per message.
In our experiments, we consider 256x256 8-bits per channel color images, that correspond to a secret message of 1.5 Mbits.



\textit{Hidden}~\cite{zhu2018hidden} and \textit{Stegastamp}~\cite{tancik2020stegastamp} also addressed the lossless case in which binary messages were hidden in a steganography image, studying the perturbations with respect to the original host image. 
{PixInWav}, on the other hand, addresses the scenario in which perturbations in the message are also acceptable, as this would be the case of visual information.
Note that images are oftentimes encoded with lossy compression algorithms whose distortions are unnoticeable for humans.
\newpage
{PixInWav} presents many points in common with \textit{Deep-Stego}~\cite{baluja2019hiding}, which hides images into images.
However, \textit{Deep-Stego} requires feeding the host and hidden images through a hiding CNN for every pair of host and hidden images. 
In contrast, {PixInWav} proposes a setup in which a single encoding of the hidden image is valid for any host audio, which greatly reduces the necessary computational requirements.

Duan et al.~\cite{duan2019reversible} proposed an approach similar to \textit{Deep-Stego}, but extended it by adopting a U-Net~\cite{ronneberger2015u} architecture and adversarial training (as in the \textit{Pix2Pix} \cite{pix2pix2017} image translation model). 
A similar U-Net is also used in {PixInWav}. 
\vspace*{-4mm}
\paragraph{Audio Steganography.}

Hiding information in audio is a relatively unexplored research area.
Early works on audio steganography relied on signal processing and audio coding \cite{santosa2005audio,atoum2013exploring,Cvejic,tekeli2017comparison,hmood2012new}, but recent advances rely on deep learning \cite{kreuk2020hide}.
Many early methods encode the hidden signal within perceptually least significant bits (LSB) of audio \cite{atoum2013exploring,Cvejic,tekeli2017comparison}.
This strategy was adopted by the only precedent, up to the author's knowledge, of hiding images in audio~\cite{hmood2012new}.
This works treated images as a generic digital signal, there was not learned visual representation or any special choice because of the visual nature of the hidden message.
This work only provided very basic qualitative results, hiding a single image in a single audio clip in a completely in silico set up.
Our experiments are much more extensive, exploring the effects of different noises, and even test it in a real world set up.

To the best of our knowledge, no previous works used deep learning for hiding images into audio. 
Only Kreuk et al.~\cite{kreuk2020hide} used neural networks for audio steganography, to send multiple speech recordings through a single host audio. 
Kreuk et al.~\cite{kreuk2020hide} noted that steganographic vision-oriented models are less suitable for audio, and propose learning a steganographic function in the frequency domain. To this end, they employ the short-time Fourier transform (STFT) and describe the challenges associated with this complex-transform (that is normally decomposed as magnitude and phase). Although many STFT-based audio models discard the phase, Kreuk et al.~\cite{kreuk2020hide} argue that discarding the phase is not practical since the decoder will be forced to also infer the phase.
To circumvent this challenge, Kreuk et al. \cite{kreuk2020hide} propose using differentiable STFT layers as part of their model. 
Our work also contributes to the discussion of which deep learning architectures are more suitable for audio steganography: (i) we propose using the short-time discrete cosine transform (STDCT, a real-transform), instead of the STFT (a complex-transform, with magnitude and phase) to avoid the above-mentioned challenges; and (ii) we propose a novel residual architecture that hides a secret image by simply adding a perceptually transparent perturbation to the host audio.

\section{Methodology}

{PixInWav} follows the classic encoder-decoder paradigm composed of two networks, which are trained end-to-end, to hide images into STDCT spectrograms.
The encoder hides an image into a host spectrogram in the shape of a (perceptually transparent) perturbation. The decoder is in charge of mapping the residualy added perturbation back into an RGB image, and it is trained to minimize the reconstruction loss with respect to the revealed (or hidden) image.
In our experiments, we apply different degrees of noise on the container audio, and assess its impact on the recovery of the secret image.


\subsection{Residual Architecture}
\label{ssec:residual}

While previous deep steganography solutions have attempted to jointly learn a representation for both the host and hidden signals, we propose to learn a representation for the hidden image only, and then add this to the host audio spectrogram.
This residual-based approach, inspired by the residual modules in ResNet~\cite{he2016deep}, makes it straightforward for the optimizer to fit an approximate identity function to the host signal, since this signal does not need to undergo a series of linear and non-linear transformations in the steganographic embedding function, as has been the case in previous work.
Note that steganography applications aim at learning an \textit{almost identity} function of the host signal, such that the hidden signal is unnoticeable during transmission.
Motivated by these ideas, we propose to simply add the encoded image to the host in a residual fashion.

\begin{figure*}[ht!]
    \centering
    \begin{subfigure}{0.49\textwidth}
        \includegraphics[width=\textwidth]{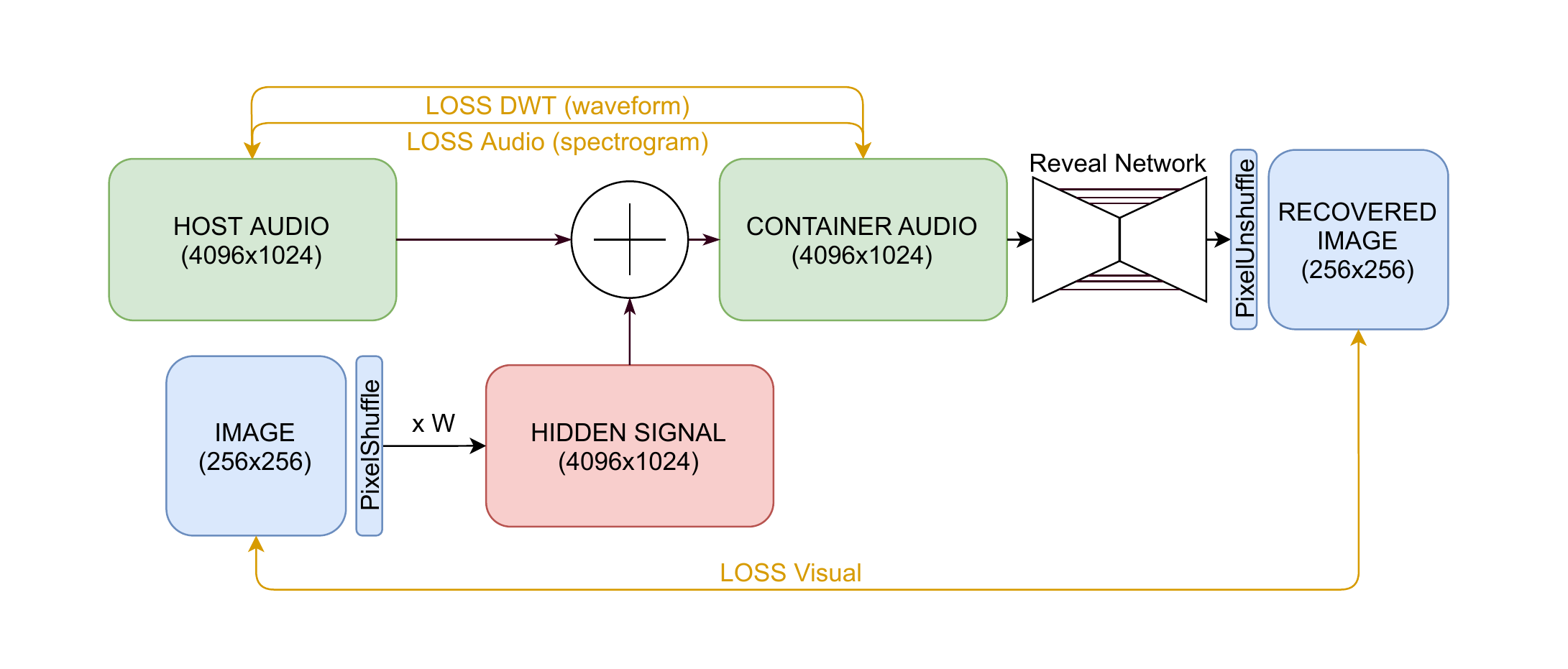}
        \caption{\small {\textbf{Res-Scale}: Residual cover-independent scaling.}}
        \label{fig:arch1}
    \end{subfigure}
    \begin{subfigure}{0.49\textwidth}
        \includegraphics[width=\textwidth]{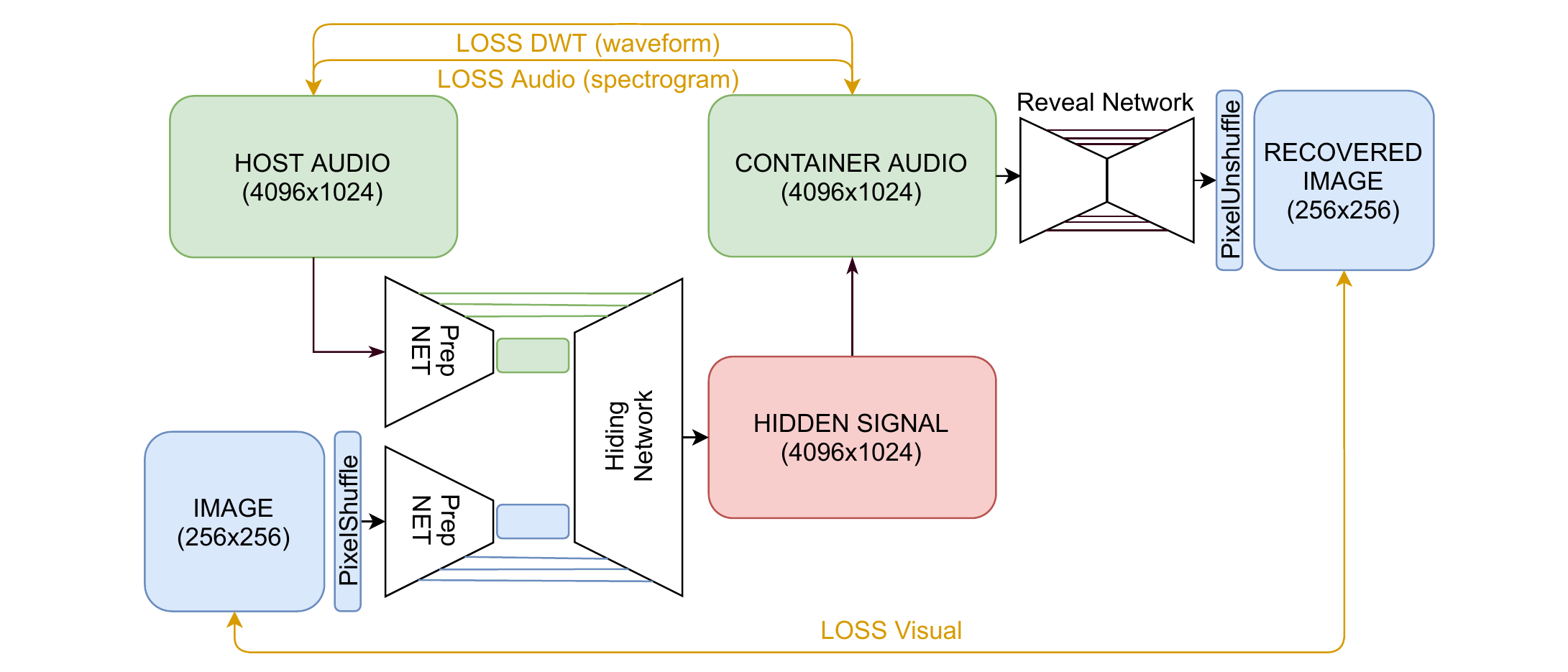}
        \caption{\small{\textbf{Plain-Dep}: Plain cover-dependent encoding.}}
        \label{fig:arch2}
    \end{subfigure}
    
    \begin{subfigure}{0.49\textwidth}
        \includegraphics[width=\textwidth]{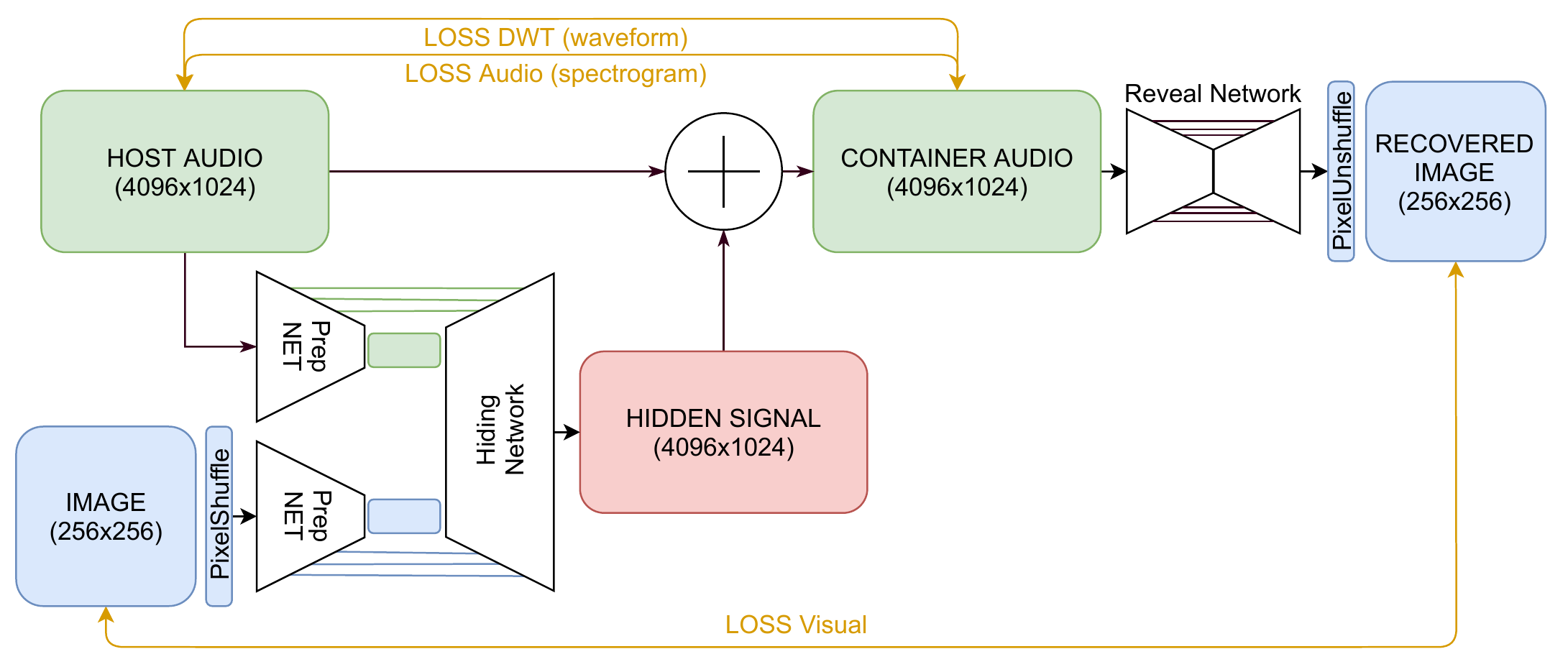}
        \caption{\small {\textbf{Res-Dep}: Residual cover-dependent encoding.}}
        \label{fig:arch3}
    \end{subfigure}
    \begin{subfigure}{0.49\textwidth}
        \includegraphics[width=\textwidth]{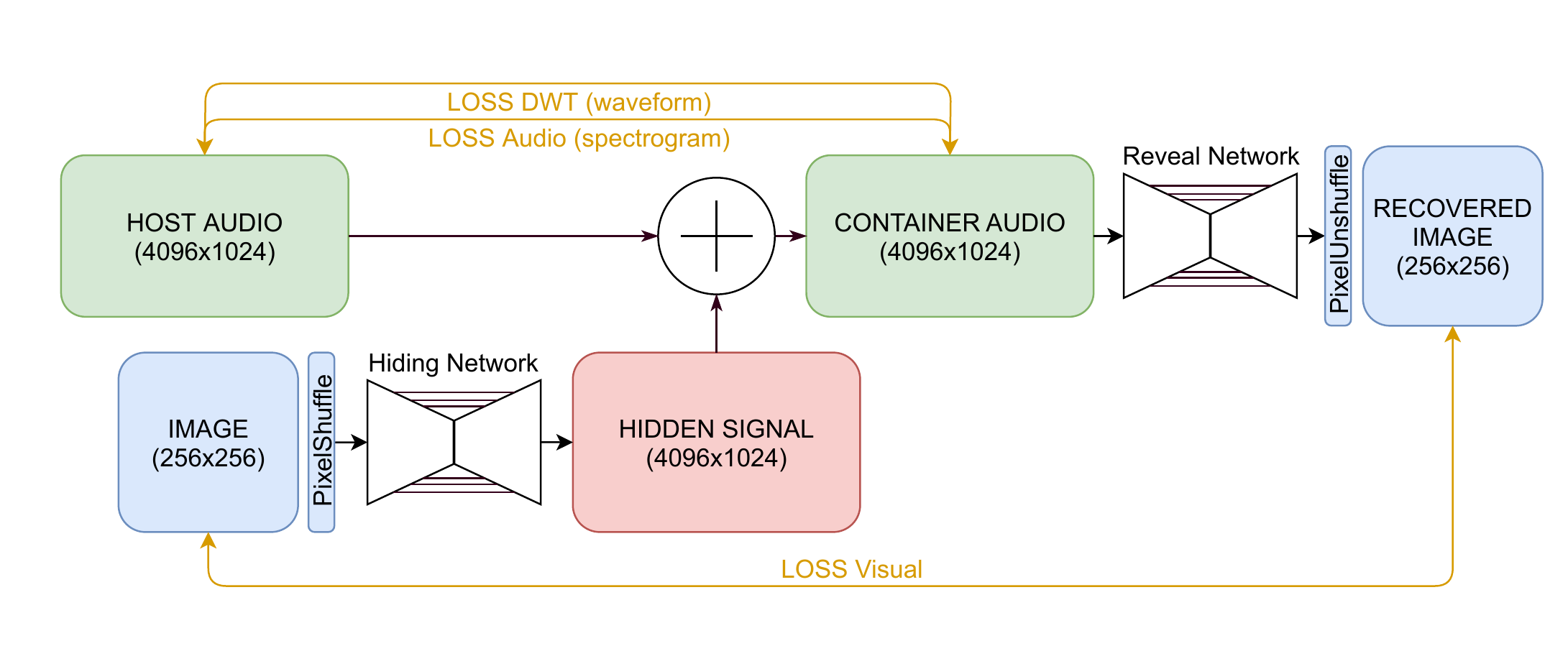}
        \caption {\small{\textbf{Res-Indep}: Residual cover-independent encoding (\textit{PixInWav}).}}
        \label{fig:arch4}
    \end{subfigure}
    \caption{\small \textbf{Neural architectures for hiding images in audio spectrograms:} The proposed {PixInWav} architecture corresponds to the \textit{Res-Indep} set up. In our ablation study, we compare against these three alternative solutions: \textit{Res-Scale}, \textit{Plain-Dep} and \textit{Res-Dep}.}
    \label{fig:architectures}
\end{figure*}

Figure \ref{fig:architectures} (d) shows the proposed {PixInWav} residual architecture, next to three other configurations we compared with in the ablation study of Section \ref{ssec:ablation}.
{PixInWav} encodes the hidden image and adds it to the STDCT-spectrogram of the host audio. 
The resulting container (stego-audio) is the signal to be transmitted and whose distortion with respect to the host audio should be perceptually unnoticeable.
At the receiver end, a decoder reconstructs the hidden image, ideally, with the minimum possible perceptual distortion.
Both the encoder and decoders are 2D fully convolutional neural networks with skip connections, based on  U-Net~\cite{ronneberger2015u}.
The encoder part, called \textit{hiding network}, contains both a contracting part (downsampling step) and an expansive part (upsampling step). The contracting part is composed by two downsampling modules, each one consisting of two $3\times3$ convolutions with stride 2 and 4, respectively. Each of the convolutional layers is followed by a batch normalization and a Leaky ReLU activation function. The expansive part is composed by two upsampling modules, each one composed of two transposed convolutional layers and a two convolutions with batch normalization. Each of these layers have a kernel size of $3\times3$ and include a Leaky ReLU activation function in between. The decoder (\textit{revealing network}) is composed of the same number of convolutional layers.

The 3-channel RGB images are augmented to four channels by appending a zero channel, subject to a $2\times 2$ pixel shuffle operation~\cite{shi2016real} to rearrange these four channels into the spatial dimensions. This operation distributes color information into the spatial domain, which makes it more straightforward for the encoding network to create residuals to be added to the spectrogram that maintain the relevant color information, and was shown necessary to obtain high-quality color reconstruction. At the output of the revealing network, the inverse pixel unshuffle operation is applied to rearrange the spatial information back into the color channels.

\subsection{Audio Representation}
\label{ssec:audio}


Next, we describe (i) why spectrograms are more suitable for residual audio steganography than waveforms, and (ii)~why we propose using STDCTs instead of relying on a more classic approach based on STFT-spectrograms.

Previous work on audio steganography relied on differentiable STFT layers to learn a steganographic function in the frequency domain \cite{kreuk2020hide}. In line with that, note that STFT-spectrograms are 2D (image-like) representations. As a result, using 2D audio representations allow for a natural way to hide (in a residual fashion) images into audio~---~since one can exploit the 2D nature of spectrograms to hide images while preserving locality.
For that reason, we discarded relying on a waveform-based model, because this would require encoding the image in a 1D signal, losing locality. 
This is particularly important under our residual setup, since the spectrogram representations we study are deterministic operations allowing perfect (inverse) reconstruction. Consequently, via employing spectrograms for residual audio steganography, we preserve the simplicity of not requiring a learnable encoder for the host signal --- but we expand our approach by allowing it to also encode locality.



As noted in previous section, using STFT-based setups can introduce difficulties related to phase reconstruction. To overcome this problem, we propose using the short-time discrete cosine transform (STDCT, a real-transform), instead of the STFT (a complex-transform, with magnitude and phase) as a simple but effective way to overcome the phase-related issues.
In short, the main difference between STFT and STDCT is the type of basis function used by each transform: the STFT uses a set of harmonically-related complex exponential basis, while the STDCT uses (real-valued) cosine basis \cite{Venkataramani}.
%
Our setup relies on the type-2 DCT.









\subsection{Loss function}

{PixInWav} is trained with a loss function allowing a trade-off between: (i)~low distortion of the host audio, and (ii)~the reconstruction quality of the hidden image. This formulation is expressed in terms of a convex combination of two reconstruction losses with a trade-off hyperparameter $\beta \in [0,1]$.

Let $s$ be the hidden image, $s'$ the revealed image, $C$ the host spectrogram and $C'$ the container spectrogram. The steganographic system is trained by minimizing the addition of the image and spectrogram reconstruction errors:
\begin{equation}
\mathcal{L}(s,s',C,C')=\beta\|s-s'\|_{1}+(1-\beta)\|C-C'\|_{2}.
\label{equation:loss-function}
\end{equation}
%
The loss function adopts a simple mean squared error (MSE) for the reconstruction of the host audio, but uses mean absolute error (MAE) to measure image reconstruction quality.

In our experiments, we also added an additional term to equation (\ref{equation:loss-function}), the soft dynamic time warping (DTW) discrepancy \cite{pmlr-v70-cuturi17a} (with $\gamma = 1)$ between the host waveform and the container waveform. The term is modulated by a constant $\lambda = 10^{-4}$ to make it comparable in magnitude to the loss in \eqref{equation:loss-function}, giving a total loss of:
\begin{equation}
    \mathcal{L}_\text{total}(s,s',C,C') = \mathcal{L}(s,s',C,C') + \lambda\; \mathbf{dtw}_\gamma(c, c'),
\end{equation}
where $c$ and $c'$ are the original and reconstructed waveforms (i.e.,~$c = \mathrm{STDCT}^{-1}(C)$).
This additional term encourages the temporal alignment between the host and container audios. For more details refer to Section \ref{ssec:dtw}. 

\section{Experiments}

\subsection{Set up}
\paragraph{Dataset}
The audio signals used in this study correspond to the FSDnoisy18K dataset~\cite{fonseca2019learning}.
This dataset contains 18,532 audio clips across 20 sound classes, depicting a large variety of sounds, such as voice, music, or noise. Since the duration of each clip is variable, we randomly select audios of approximately 1.5 seconds at 44,100Hz. We computed the STDCT transform with a frame length $2^{12}$ and a hop size of $2^{6}-2$. These hyperparameters where chosen to obtain a spectrogram with width and height being powers of 2, which allows for efficient computations. 


RGB images were sampled from the ImageNet (ILSVRC2012) dataset~\cite{russakovsky2015imagenet}. 10,000 randomly sampled images were used to train PixInWav, while validation results are reported over a non-overlapping partition of 900 images. Each RGB image was resized, cropped and normalized, resulting in a $256 \times 256 \times 3$ image, and paired with a randomly selected sound from the audio dataset.

\paragraph{Training details}

The model was trained with the Adam optimizer at a learning rate ($lr$) of $0.01$ and a batch size of 1. Additional experiments with $lr=0.1,0.001$ did not converge. 
Leaky-ReLU are set with $\alpha=0.8$.
The revealed image at the output is clipped in the range of $[0,1]$ and denormalized back to the range of RGB values: $[0,255]$. 

\paragraph{Evaluation metrics}
The inclusion of the image into the audio signal introduces a distortion, which we measure with the signal-to-noise-ratio (SNR) over the waveform --- where the noise corresponds to the difference between the host and container audios.
We adopted SNR as audio quality metric because it is widely used among related works~\cite{santosa2005audio,kreuk2020hide}.


Analogously, the image also suffers a distortion as a result of its encoding in the hiding network, mixing with the audio and decoding with the reveal network.
We measure the visual distortion with the Structural Similarity Index (SSIM)~\cite{wang2004image}, a  metric that takes into consideration the perceptual properties of the human visual system.
As a complement, we also provide results in peak signal-to-noise-ratio (PSNR), to allow contrasting our values with the literature~\cite{baluja2019hiding}.

\subsection{Results}
\paragraph{Trade-off between audio-image distortions.}
\label{ssec:beta}
{PixInWav} was trained with different distortion trade-offs between the host audio and hidden image, governed by the $\beta$ parameter of the loss function.
The impact of $\beta$ was studied both from a quantitative (Figure~\ref{fig:beta}) and qualitative (Table~\ref{tab:beta-span}) perspective. Audio samples are provided in the supplementary material.


\begin{figure}
\begin{center}
   \pgfplotsset{compat = 1.3}
\begin{center}
\begin{minipage}{0.95\linewidth}
\resizebox{\linewidth}{!}{\begin{tikzpicture}[/pgfplots/width=0.95\linewidth, /pgfplots/height=\linewidth, /pgfplots/legend pos=south east]
\begin{axis}[
     	ymin=-16, ymax=20,
        xmin=0.05, xmax=0.9,
        ylabel=Audio SNR,
        xlabel=$\beta$,
		font=\large,
        grid=both,
		grid style=dotted,
        xmode=linear,
        ytick={-10, 0, 10, 20},
		yticklabels={-10, 0, 10, 20},
        xtick={0.1, 0.25, 0.5, 0.75, 0.9},
	    xticklabels={0.1, 0.25, 0.5, 0.75, 0.9},
        enlarge x limits=0.00,
        legend style={at={(2.16,1.1)}, anchor=north, /tikz/every even column/.append style={column sep=3.6mm}},
        legend columns = -1,
        ]
        \addlegendentry{With DTW loss}
        \addplot[smooth,mark=x,blue] 
          coordinates{
            (0.05,18.41)
            (0.1,15.72)
            (0.25,12.89) 
            (0.5,9.01)
            (0.75,7.88)
            (0.9,6.23)
        }; \label{audioSNRyesDTW}   
        
        \addlegendentry{Without DTW loss}
        \addplot[smooth,mark=x,blue, dashed] 
          coordinates{
            (0.05,5.061)
            (0.1,5.061)
            (0.25,-3.425) 
            (0.5,-9.496)
            (0.75,-12.084)
            (0.9,-15.306)
        }; \label{audioSNRnoDTW}
    \end{axis}

\begin{axis}[
    	ymin=0.89, ymax=1.0,
        xmin=0.1, xmax=0.9,
        ylabel=image SSIM,
        xlabel=$\beta$,
		font=\large,
        grid=both,
		grid style=dotted,
        xmode=linear,
        ytick={0.90, 0.95, 1.00},
		yticklabels={0.90, 0.95, 1.00},
        xtick={0.1, 0.25, 0.5, 0.75, 0.9},
	    xticklabels={0.1, 0.25, 0.5, 0.75, 0.9},
        enlarge x limits=0.00,
        ylabel shift=0.0cm,
        xshift=8.5cm
        ]
        
        \addplot[smooth,mark=x,blue] 
          coordinates{
            (0.05,0.921)
            (0.1,0.930)
            (0.25,0.936) 
            (0.5,0.937)
            (0.75,0.957)
            (0.9,0.962)
        }; \label{audioSNRyesDTW} 
		
		\addplot[smooth,mark=x,blue, dashed]  
         coordinates{
            (0.05,0.9283)
            (0.25,0.948) 
            (0.5,0.959)
            (0.75,0.967)
            (0.9,0.972)};

    \end{axis}
\end{tikzpicture}}
\end{minipage}
\end{center}
\end{center}
   \caption{Quality trade-off between the host audio (left) and the hidden image (right). The $x$-axis represents the hyperparameter $\beta$ that controls the distortion between the two. The reported values correspond to the $8^{th}$ training epoch.}
\label{fig:beta}
\end{figure}
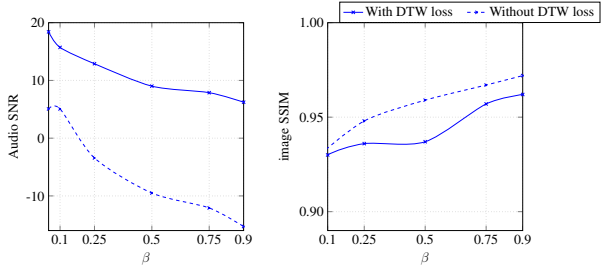

\begin{table}
\begin{center}
\resizebox{1.0\linewidth}{!}{
\begin{tabular}{cccc}
 $\beta=0.05$ & $\beta=0.1$    & $\beta=0.5$ & $\beta=0.9$\\
\includegraphics[width=.29\linewidth]{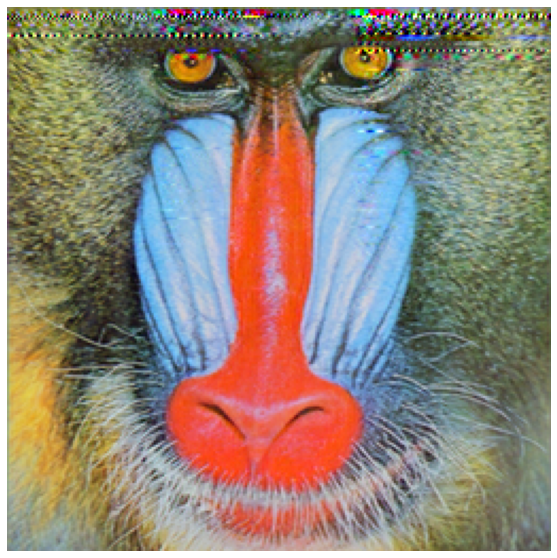} &
\includegraphics[width=.29\linewidth]{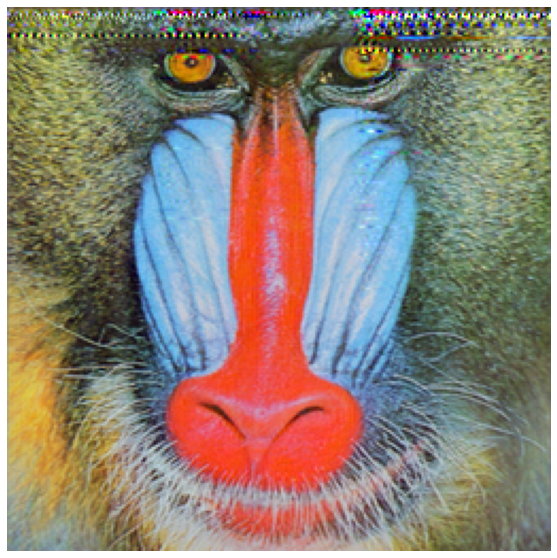}   &      \includegraphics[width=.29\columnwidth]{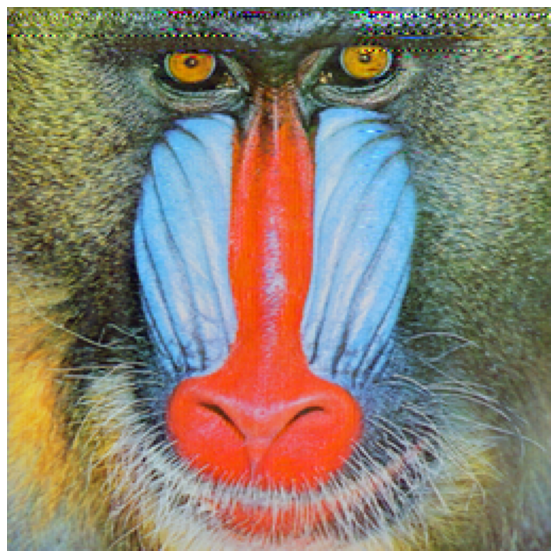} &
\includegraphics[width=.29\columnwidth]{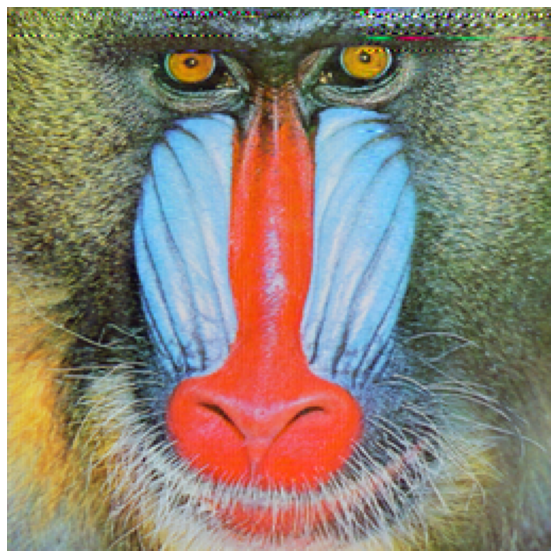}
\\ 
\includegraphics[width=.29\linewidth]{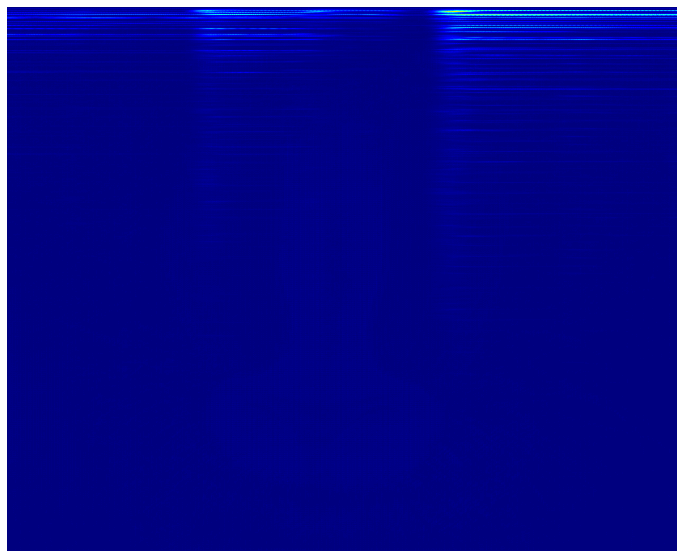} &
\includegraphics[width=.29\linewidth]{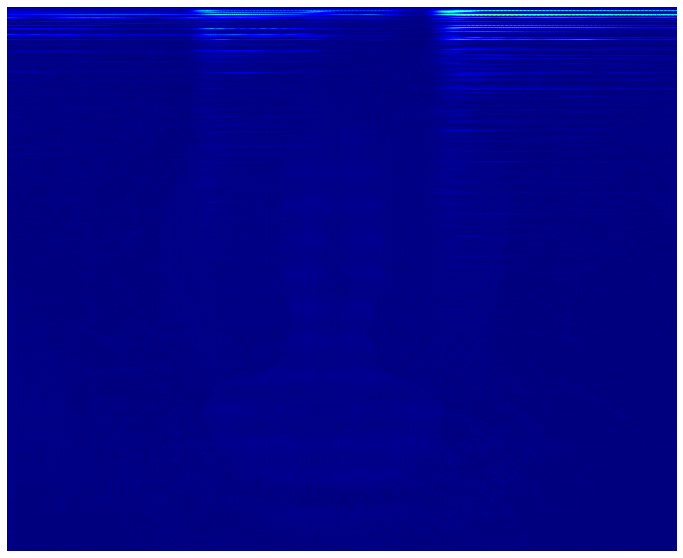}   &      \includegraphics[width=.29\columnwidth]{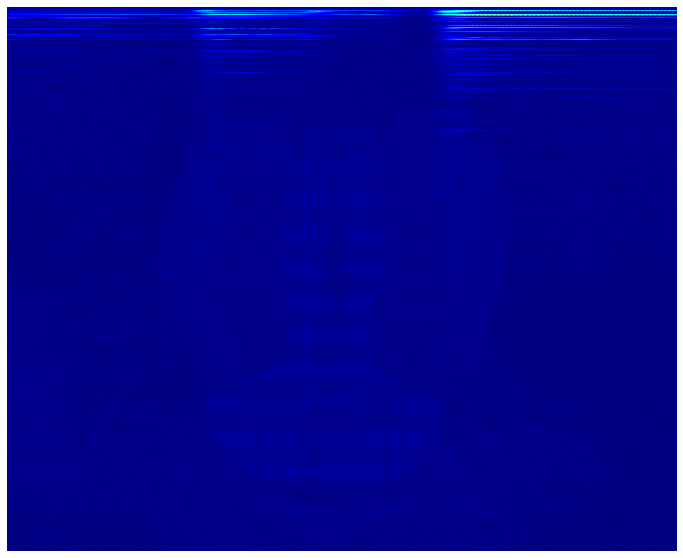} &
\includegraphics[width=.29\columnwidth]{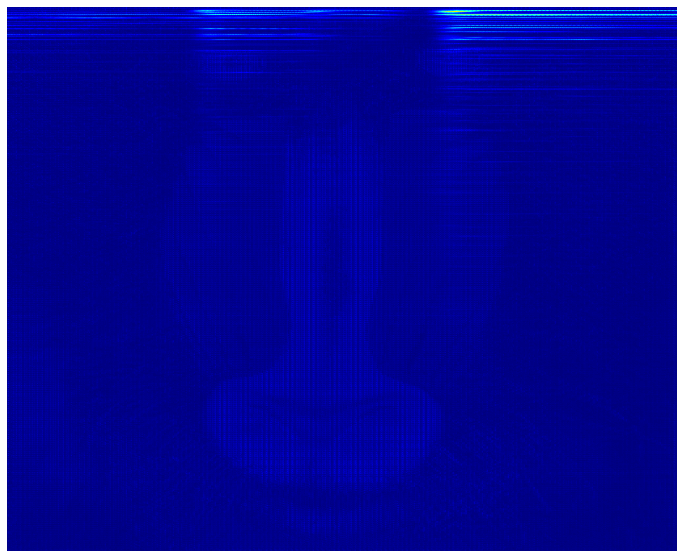}
\end{tabular}}
\end{center}
\caption{\textbf{Effect of $\beta$ on the image and spectrogram:} The degradation over the image quality is hardly noticeable, but the silhouette of the baboon becomes visible in the spectrogram for the higher $\beta$.}
\label{tab:beta-span}
\end{table}

As expected, lower $\beta$ values preserve better the quality of the audio, while higher ones allow a better recovery of the hidden image, at the expense of a reduction of audio SNR.
These obtained results demonstrate that \textit{PixInWav} can meet perceptually acceptable quality standards for both the host audio and the hidden image.
As a reference, in the speech domain \cite{ICSIspeechFAQ}, it is considered that listeners will barely notice any distortion when the SNR is above 20~dB, and intelligibility is still reasonable at 0~dB SNR (speech energy and noise energy the same).
In the remainder of our experiments, we set a $\beta$ of $0.05$, which corresponds to an average audio SNR of $18.26$~dB and an average SSIM of $0.921$ for a model trained during 8 epochs.



\begin{figure}
\centering
\begin{subfigure}{0.45\columnwidth}
    \centering
    \includegraphics[width=\columnwidth]{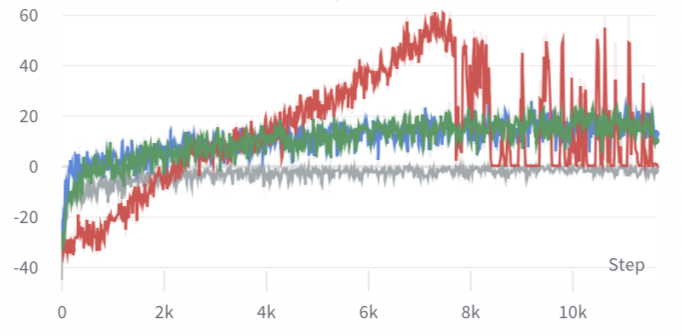}
    \caption[short]{Audio SNR.}
\end{subfigure}
\begin{subfigure}{0.45\columnwidth}
    \centering
    \includegraphics[width=\columnwidth]{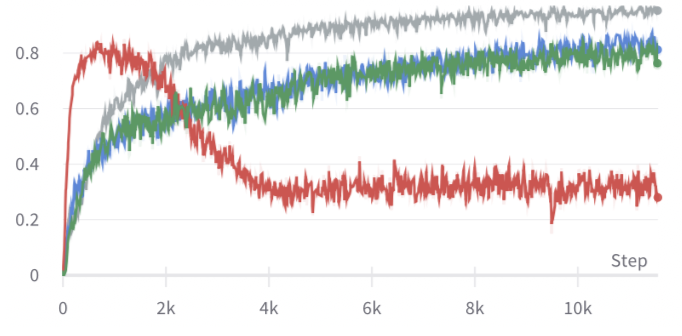}
    \caption[short]{Image SSIM.}
\end{subfigure}\\

\caption[short]{Validation curves along training iterations for the architectures \textcolor{BrickRed}{Res-Scale (in red)}, \textcolor{Gray}{Plain-Dep (in grey)}, \textcolor{NavyBlue}{Res-Dep (in blue)},  and \textcolor{OliveGreen}{Res-Indep (in green)}. \textcolor{NavyBlue}{Res-Dep} and \textcolor{OliveGreen}{Res-Indep} present a similar behaviour. \textcolor{Gray}{Plain-Dep} fails in achieving an acceptable audio SNR, while \textcolor{BrickRed}{Res-Scale} becomes unstable for the audio, and drops its performance for images after being trained for 1,000 iterations.}
\label{fig:losses}
\end{figure}

\vspace*{-2mm}
\paragraph{Dynamic time warping loss on audio.}
\label{ssec:dtw}
Figure \ref{fig:beta} also plots the corresponding SNR and SSIM values if the DTW loss term was not included.
The results show the importance of this term, as removing it drops the audio SNR more than 10 dB, moving below the 0 dB case for most tested $\beta$.
On the other hand, the DTW loss term applied over the audio signal actually introduces only a small distortion over the images,
unnoticeable for a human.
\vspace*{-2mm}
\paragraph{Ablation study.}
\label{ssec:ablation}
The \textit{Res-Indep} architecture for PixInWav proposed in Section \ref{ssec:residual} was compared with the other three approaches depicted in Figure \ref{fig:architectures}: \textit{Res-Scale}, \textit{Plain-Dep} and \textit{Res-Dep}. \textit{Plain-Dep} is used to compare our residual approach with a classic feedforward encoder-decoder architecture, and \textit{Res-Dep} shows the effect of conditioning the hiding network on both the image and the host audio. \textit{Res-Scale} is included to check if the hiding network is not simply uniformly encoding the image signal in the low-order bits of the host signal. We compared the four solutions based on the quantitative (Table \ref{tab:ablation}) and qualitative results (Table \ref{tab:ablation-qualitative}), as well as the convergence of their validation loss curves (Figure \ref{fig:losses}).

\begin{table}
\centering

\begin{tabular}{@{}lccc@{}}
\toprule
                     & \multicolumn{1}{c}{Audio} & \multicolumn{2}{c}{Image}\\
Architecture        & SNR $\uparrow$ & SSIM $\uparrow$ & PSNR $\uparrow$\\
\midrule
Res-Scale      &    $54.19$       & $0.294$       & $12.49$\\
Plain-Dep       &     $-1.09$     & $0.926$       & $29.20$ \\
Res-Dep          &     $17.02$      & $0.814$       & $21.34$\\
Res-Indep         &    $17.45$        & $0.832$       & $22.67$\\
\bottomrule
\end{tabular}
\caption{\textbf{Ablation study:} Audio and image quality metrics for a fixed $\beta=0.05$ with \textit{PixInWav} and the three considered baselines. Results are after one training epoch. The supplementary material includes results for more epochs.}
\label{tab:ablation}
\end{table}

\begin{table}
\begin{center}
\resizebox{1.0\linewidth}{!}{
\begin{tabular}{cccc}
 Res-Scale & Plain-Dep    & Res-Dep  & Res-Indep \\
\includegraphics[trim={2cm 2cm 2cm 1.75cm},width=.29\linewidth]{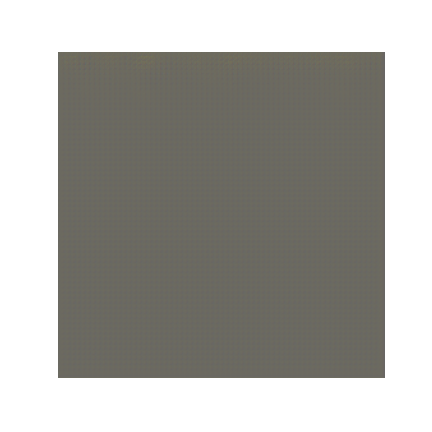} &
\includegraphics[trim={2cm 2cm 2cm 1.75cm}, width=.29\linewidth]{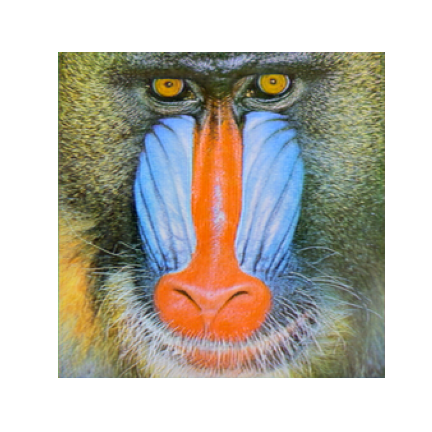}   & \includegraphics[trim={2cm 2cm 2cm 1.75cm},width=.29\linewidth]{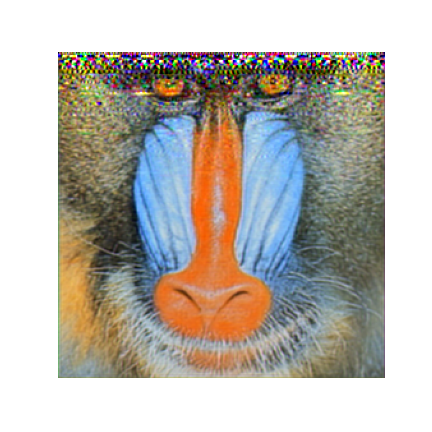}   &      \includegraphics[trim={2cm 2cm 2cm 1.75cm},width=.29\columnwidth]{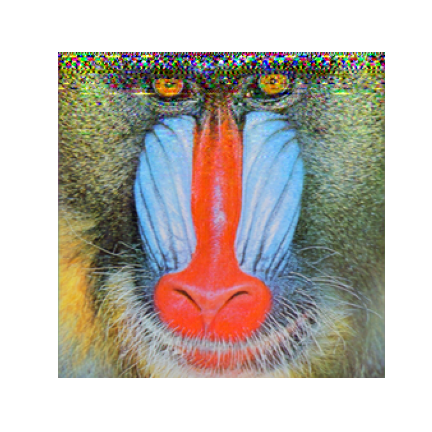} \\ 
\includegraphics[trim={2cm 2cm 2cm 1.75cm},width=.29\linewidth]{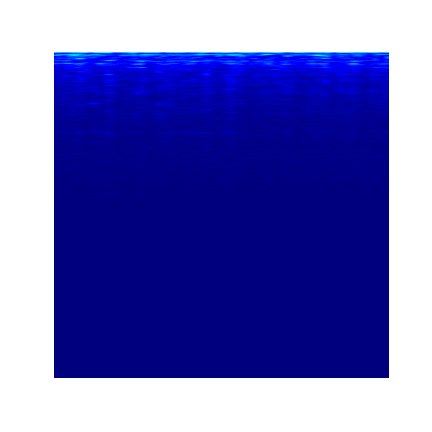} &
\includegraphics[trim={2cm 2cm 2cm 1.75cm},width=.29\linewidth]{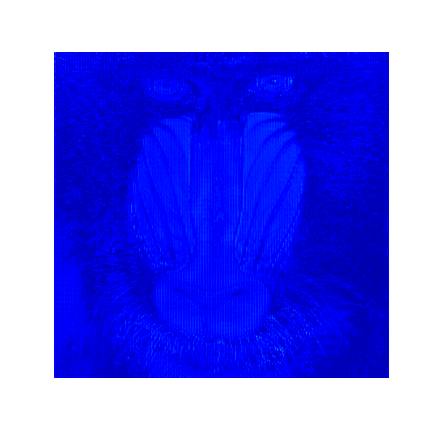}  & \includegraphics[trim={2cm 2cm 2cm 1.75cm},width=.29\linewidth]{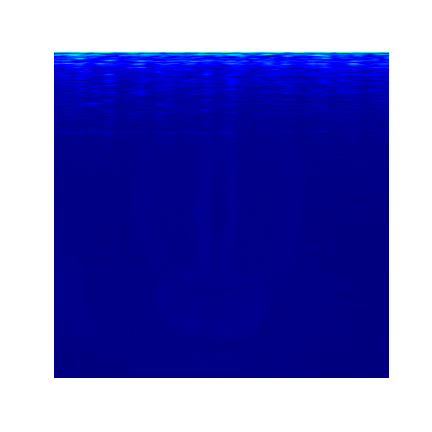}   &      \includegraphics[trim={2cm 2cm 2cm 1.75cm},width=.29\columnwidth]{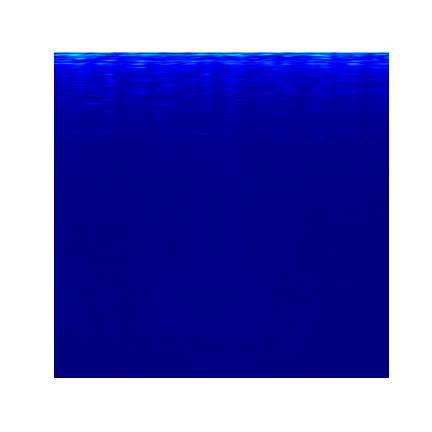} 
\end{tabular}}
\end{center}
\caption{\textbf{Qualitative comparison between architectures:} \textit{Res-Scale} fails in transmitting the image and \textit{Plain-Dep} allows a visible interference of the image over the spectrogram. On the other hand,  \textit{Res-Dep} and \textit{Res-Indep} provide similar qualitative results, while the later reproduces more similar colors to the original image.}
\label{tab:ablation-qualitative}
\end{table}

The best quantitative and qualitative results are obtained with the \textit{Res-Dep} and \textit{Res-Indep} solutions, both achieving similar performance and learning behaviour.
We hypothesize that the hiding network in \textit{Res-Dep} learns to reduce visual information in those frequencies where audio is present. 
The distribution of these frequencies can be learned as a prior with \textit{Res-Indep} so that, in the end, the performance is similar to \textit{Res-Dep}.
This may explain the minimal differences in terms of SNR or SSIM between both configurations.

\textit{Res-Scale} prevents the image from being recovered, while \textit{Plain-Dep} fails in transmitting the audio.
As a consequence, these two options are discarded.


We adopt \textit{Res-Indep} as the reference architecture in further experiments, because its independent representation with respect to the host makes it especially suitable for steganography applications.
A unique representation for the hidden image can be embedded in any audio signal, which is a much more scalable solution than computing a specific transformation for each audio snippet.
Table \ref{tab:qualitative} provides qualitative results of the both the original host and container STDCT spectrograms, together with their matching original and recovered images.

\begin{table*}
\begin{center}
\resizebox{1.0\linewidth}{!}{
\begin{tabular}{cccc}
 \multicolumn{2}{c}{Originals}  &   Container      &   Recovered  \\
 Host    & Hidden            & Host + Hidden & Hidden \\
\includegraphics[trim={1.5cm 1.5cm 1.5cm 1.5cm},width=.2\linewidth]{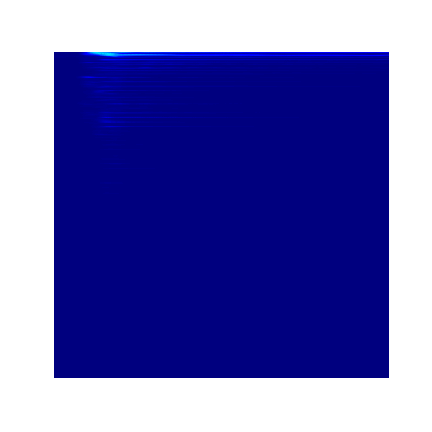}   & \includegraphics[trim={1.5cm 1.5cm 1.5cm 1.5cm},width=.2\linewidth]{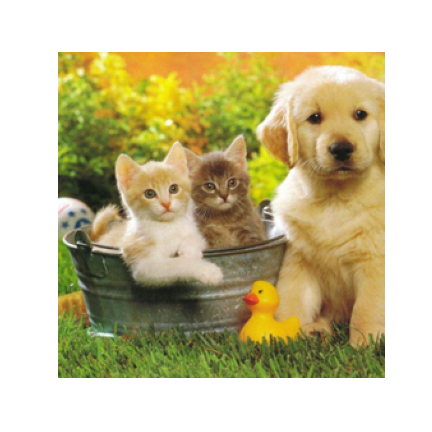}   & \includegraphics[trim={1.5cm 1.5cm 1.5cm 1.5cm},width=.2\linewidth]{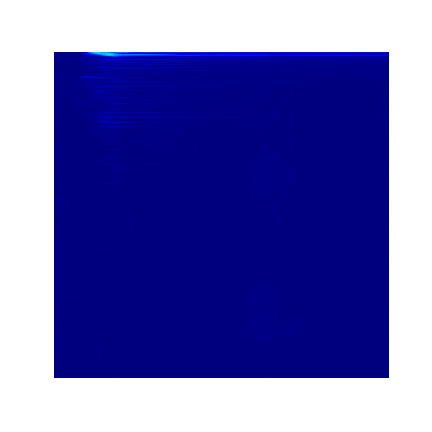} & \includegraphics[trim={1.5cm 1.5cm 1.5cm 1.5cm},width=.2\linewidth]{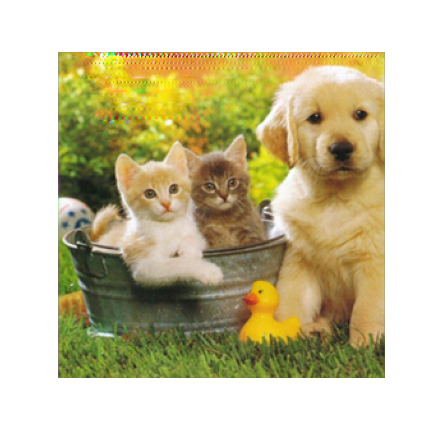} \\ 
\includegraphics[trim={1.5cm 1.5cm 1.5cm 1.5cm},width=.2\linewidth]{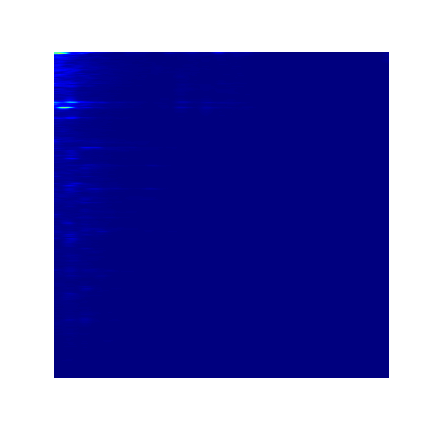}   & \includegraphics[trim={1.5cm 1.5cm 1.5cm 1.5cm},width=.2\linewidth]{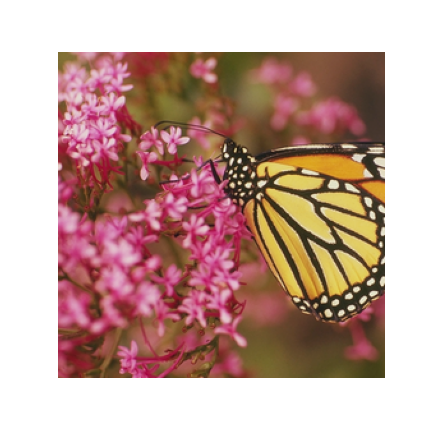}   & \includegraphics[trim={1.5cm 1.5cm 1.5cm 1.5cm},width=.2\linewidth]{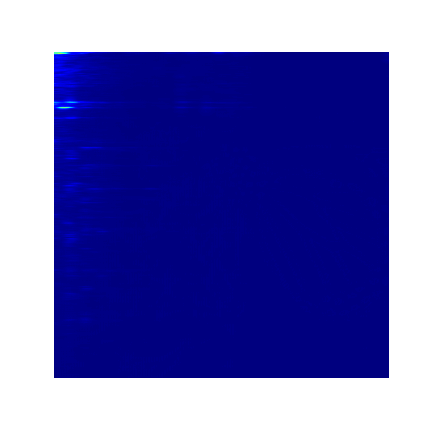} & \includegraphics[trim={1.5cm 1.5cm 1.5cm 1.5cm},width=.2\linewidth]{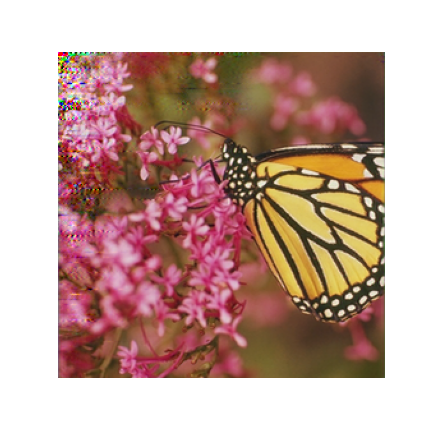} \\ 
\includegraphics[trim={1.5cm 1.5cm 1.5cm 1.5cm},width=.2\linewidth]{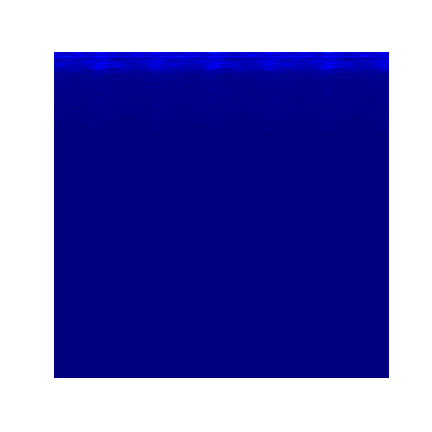}   & \includegraphics[trim={1.5cm 1.5cm 1.5cm 1.5cm},width=.2\linewidth]{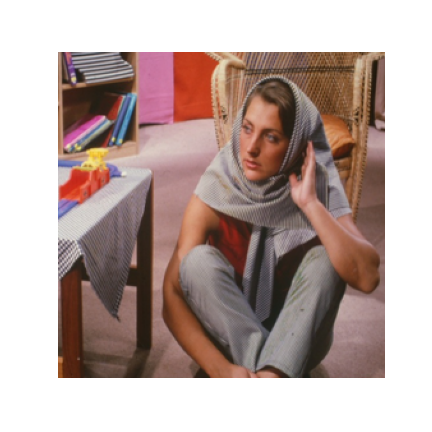}   & \includegraphics[trim={1.5cm 1.5cm 1.5cm 1.5cm},width=.2\linewidth]{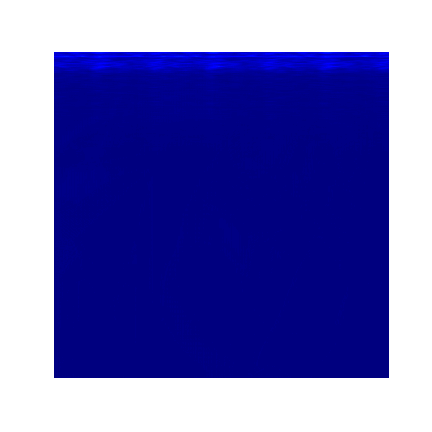} & \includegraphics[trim={1.5cm 1.5cm 1.5cm 1.5cm},width=.2\linewidth]{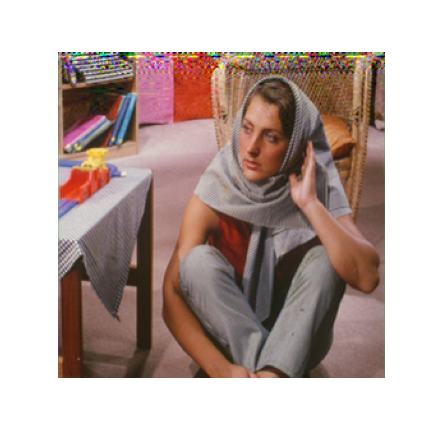} \\ 
\includegraphics[trim={1.5cm 1.5cm 1.5cm 1.5cm},width=.2\linewidth]{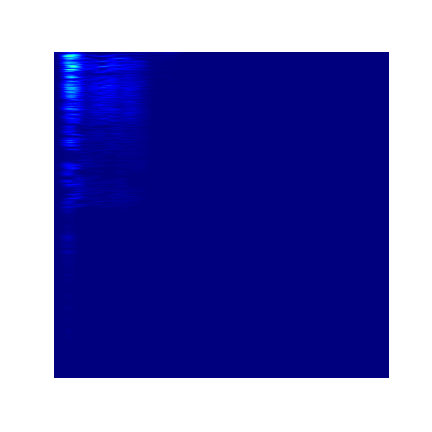}   & \includegraphics[trim={1.5cm 1.5cm 1.5cm 1.5cm},width=.2\linewidth]{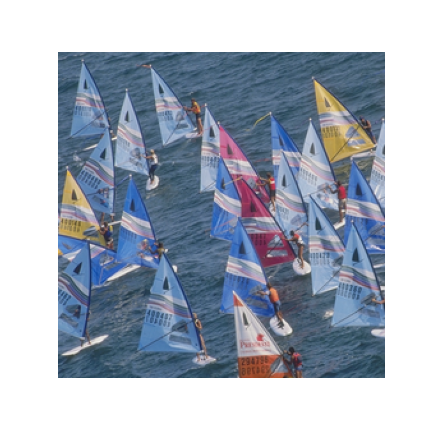}   & \includegraphics[trim={1.5cm 1.5cm 1.5cm 1.5cm},width=.2\linewidth]{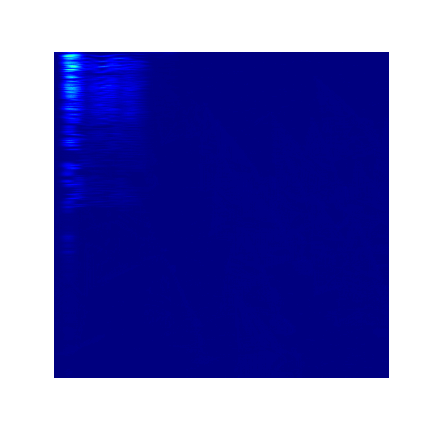} & \includegraphics[trim={1.5cm 1.5cm 1.5cm 1.5cm},width=.2\linewidth]{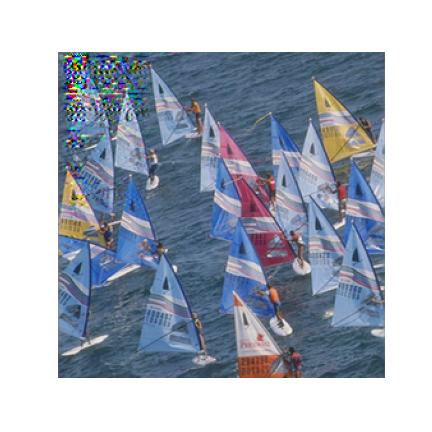} \\ 
\includegraphics[trim={1.5cm 1.5cm 1.5cm 1.5cm},width=.2\linewidth]{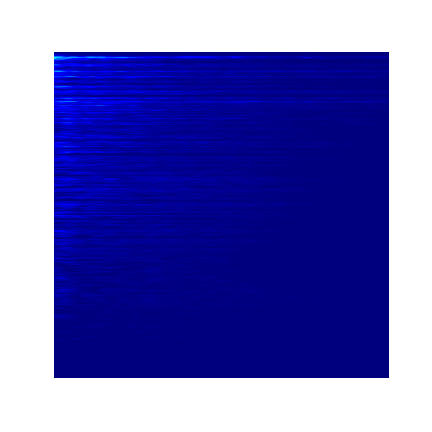}   & \includegraphics[trim={1.5cm 1.5cm 1.5cm 1.5cm},width=.2\linewidth]{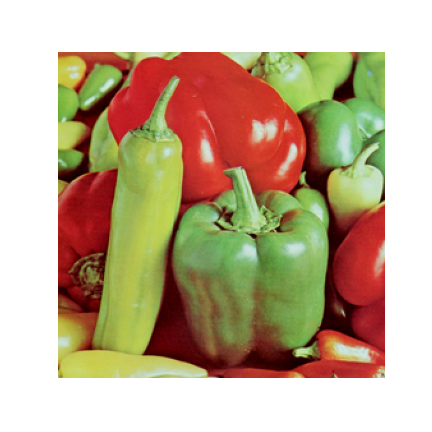}   & \includegraphics[trim={1.5cm 1.5cm 1.5cm 1.5cm},width=.2\linewidth]{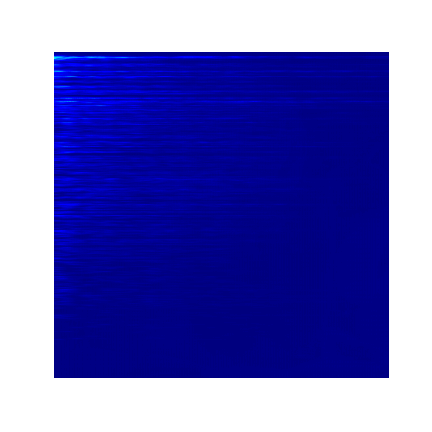} & \includegraphics[trim={1.5cm 1.5cm 1.5cm 1.5cm},width=.2\linewidth]{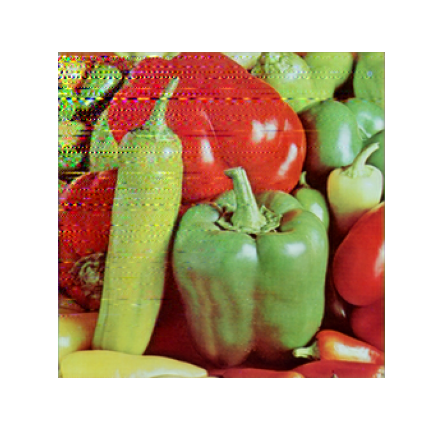} \\ 
\end{tabular}}
\end{center}
\caption{\textbf{Diverse qualitative results:} The first two columns show the original host spectrogram and hidden image, which are combined in the container at the third column. The last column presents the recovered hidden image. A detailed analysis of this last column allows identifying a footprint of one modality into the other, especially of the high power locations of the STDCT spectrogram over the image pixels. }
\label{tab:qualitative}
\end{table*}

\paragraph{Robustness to distortions}
In a real world scenario, the audio signal may be exposed to distortions resulting from the channel response or compression algorithms.
In our work, we simulate these conditions by adding additive white Gaussian noise (AWGN) or speckle noise, as proposed in~\cite{kreuk2020hide} and with the source code provided by the authors.
The power of both noises is governed by a hyperparameter $\sigma$, which corresponds to the norm of the added noise.

In a first experiment, we study the impact of training \textit{PixInWav} by adding some AWGN or speckle noise, and later test it in a noise-free environment.
The quantitative results presented in Table \ref{tab:noise-notatval} do not indicate any significant change of performance with respect to the baseline case of training PixInWav without synthetic noise.

\begin{table}
\begin{tabular}{@{}ccccc@{}}
\toprule
\multicolumn{2}{c}{}                & \multicolumn{1}{c}{Audio} & \multicolumn{2}{c}{Image}\\
  Noise & $\sigma$      & SNR $\uparrow$ & SSIM $\uparrow$ & PSNR $\uparrow$\\
\midrule
 -       &     -     & $18.41$            & $0.92$      & $27.37$\\
AWGN     & $0.1$      & $18.80$             & $0.86$      & $27.30$\\
Speckle & $0.1$     & $15.56$            & $0.92$      & $27.96$\\

\bottomrule
\end{tabular}
\caption{\textbf{Study of the impact of adding synthetic noise at training time:} Compared to the baseline case of not training with any noise (1st row), the introduction of some AWGN (2nd row) or speckle (3rd row) noise does not produce a significant impact when PixInWav is operated in a noise-free environment.}
\label{tab:noise-notatval}
\end{table}

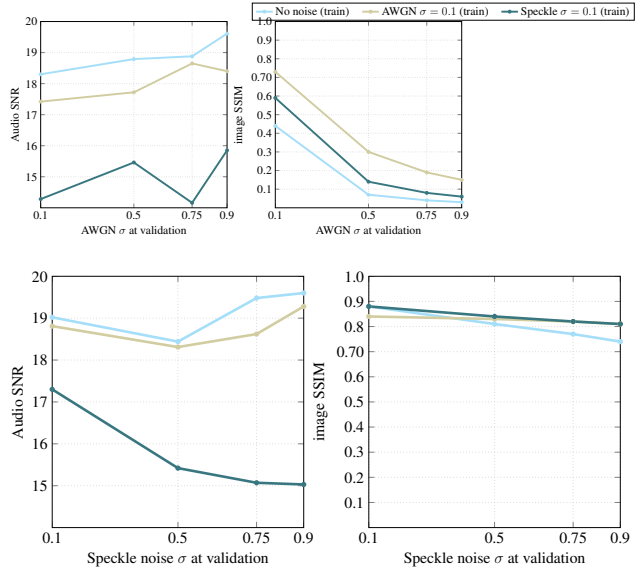
\begin{figure}
\begin{center}
   \pgfplotsset{compat = 1.3}
\begin{center}
\begin{minipage}{\linewidth}
\resizebox{\linewidth}{!}{\begin{tikzpicture}[/pgfplots/width=\linewidth, /pgfplots/height=\linewidth, /pgfplots/legend pos=south east]
\begin{axis}[
    	ymin=14, ymax=20,
        xmin=0.1, xmax=0.9,
        ylabel=Audio SNR,
        xlabel=AWGN $\sigma$ at validation,
		font=\large,
        grid=both,
		grid style=dotted,
        xmode=linear,
        ytick={15, 16, 17, 18, 19, 20},
		yticklabels={15, 16, 17, 18, 19, 20},
        xtick={0.1, 0.5, 0.75, 0.9},
	    xticklabels={0.1, 0.5, 0.75, 0.9},
        enlarge x limits=0.00,
        legend style={at={(2.16,1.1)}, anchor=north, /tikz/every even column/.append style={column sep=3.6mm}},
        legend columns = -1,
        ]
        
        \addlegendentry{No noise (train)}
		\addplot[color=c_b0,mark=*, mark size=1, line width=2] coordinates{(0.1,18.30)(0.5,18.79)(0.75,18.88)(0.9,19.61)};
		
        \addlegendentry{AWGN $\sigma=0.1$ (train)}
		\addplot[color=c_b0.1,mark=*, mark size=1, line width=2] coordinates{(0.1,17.42)(0.5,17.72)(0.75,18.65)(0.9,18.40)};
		
        \addlegendentry{Speckle $\sigma=0.1$ (train)}
		\addplot[color=c_b0.5,mark=*, mark size=1, line width=2] coordinates{(0.1,14.28)(0.5,15.46)(0.75,14.16)(0.9,15.85)};
    \end{axis}

\begin{axis}[
    	ymin=0.0, ymax=1.0,
        xmin=0.1, xmax=0.9,
        ylabel=image SSIM,
        xlabel=AWGN $\sigma$ at validation,
		font=\large,
        grid=both,
		grid style=dotted,
        xmode=linear,
        ytick={0.1, 0.2, 0.3, 0.4, 0.5, 0.6, 0.7, 0.8, 0.9, 1.0},
		yticklabels={0.1, 0.2, 0.3, 0.4, 0.5, 0.6, 0.70, 0.8, 0.9, 1.0},
        xtick={0.1, 0.5, 0.75, 0.9},
	    xticklabels={0.1, 0.5, 0.75, 0.9},
        enlarge x limits=0.00,
        ylabel shift=0.0cm,
        xshift=8.5cm
        ]
        
		\addplot[color=c_b0,mark=*, mark size=1, line width=2] coordinates{(0.1,0.44)(0.5,0.07)(0.75,0.04)(0.9,0.03)};

    	\addplot[color=c_b0.1,mark=*, mark size=1, line width=2] coordinates{(0.1,0.73)(0.5,0.30)(0.75,0.19)(0.9,0.15)};

		\addplot[color=c_b0.5,mark=*, mark size=1, line width=2] coordinates{(0.1,0.59)(0.5,0.14)(0.75,0.08)(0.9,0.06)};

    \end{axis}
\end{tikzpicture}}
\end{minipage}
\end{center}

\pgfplotsset{compat = 1.3}
\begin{center}
\begin{minipage}{\linewidth}
\resizebox{\linewidth}{!}{\begin{tikzpicture}[/pgfplots/width=\linewidth, /pgfplots/height=\linewidth, /pgfplots/legend pos=south east]
\begin{axis}[
    	ymin=14, ymax=20,
        xmin=0.1, xmax=0.9,
        ylabel=Audio SNR,
        xlabel=Speckle noise $\sigma$ at validation,
		font=\large,
        grid=both,
		grid style=dotted,
        xmode=linear,
        ytick={15, 16, 17, 18, 19, 20},
		yticklabels={15, 16, 17, 18, 19, 20},
        xtick={0.1, 0.5, 0.75, 0.9},
	    xticklabels={0.1, 0.5, 0.75, 0.9},
        enlarge x limits=0.00,
        ]
		\addplot[color=c_b0, mark=*, mark size=1, line width=2] coordinates{(0.1,19.02)(0.5,18.44)(0.75,19.48)(0.9,19.60)};
		\addplot[color=c_b0.1, mark=*, mark size=1, line width=2] coordinates{(0.1,18.81)(0.5,18.31)(0.75,18.62)(0.9,19.28)};
		\addplot[color=c_b0.5, mark=*, mark size=1, line width=2] coordinates{(0.1,17.30)(0.5,15.42)(0.75,15.07)(0.9,15.03)};
    \end{axis}

\begin{axis}[
    	ymin=0.0, ymax=1.0,
        xmin=0.1, xmax=0.9,
        ylabel=image SSIM,
        xlabel=Speckle noise $\sigma$ at validation,
		font=\large,
        grid=both,
		grid style=dotted,
        xmode=linear,
        ytick={0.1, 0.2, 0.3, 0.4, 0.5, 0.6, 0.7, 0.8, 0.9, 1.0},
		yticklabels={0.1, 0.2, 0.3, 0.4, 0.5, 0.6, 0.70, 0.8, 0.9, 1.0},
        xtick={0.1, 0.5, 0.75, 0.9},
	    xticklabels={0.1, 0.5, 0.75, 0.9},
        enlarge x limits=0.00,
        ylabel shift=0.0cm,
        xshift=8.5cm
        ]
        
		\addplot[color=c_b0, mark=*, mark size=1, line width=2] coordinates{(0.1,0.88)(0.5,0.81)(0.75,0.77)(0.9,0.74)};
    	\addplot[color=c_b0.1, mark=*, mark size=1, line width=2] coordinates{(0.1,0.84)(0.5,0.83)(0.75,0.82)(0.9,0.81)};
		\addplot[color=c_b0.5, mark=*, mark size=1, line width=2]
		coordinates{(0.1,0.88)(0.5,0.84)(0.75,0.82)(0.9,0.81)};
    \end{axis}
\end{tikzpicture}}
\end{minipage}
\end{center}
\end{center}
   \caption{\textbf{Effect of Gaussian and speckle noise during training:} The top row measures the effect of AWGN on models trained with or without noise, while the bottom row presents the analogous experiment with speckle noise. In general, performance decreases when increasing the noise. However, the audio SNR of PixInWav seems to be more damaged by speckle noise when trained with speckle noise, than without noise or AWGN.}
\label{fig:noise-val}
\end{figure}

\begin{figure}
\centering
\begin{subfigure}{\columnwidth}
    \centering
    \includegraphics[width=.19\columnwidth]{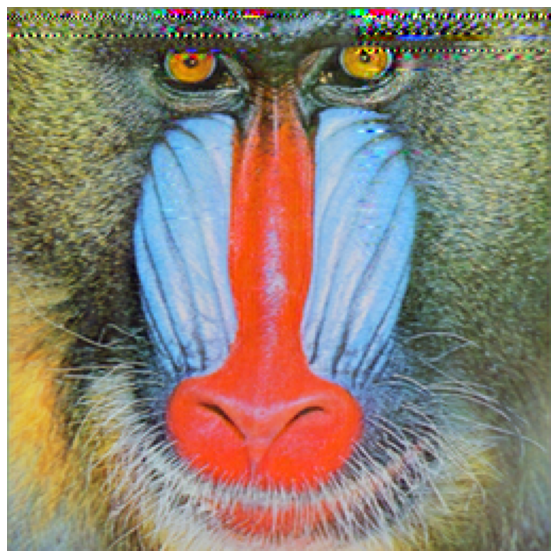}
    \includegraphics[width=.19\columnwidth]{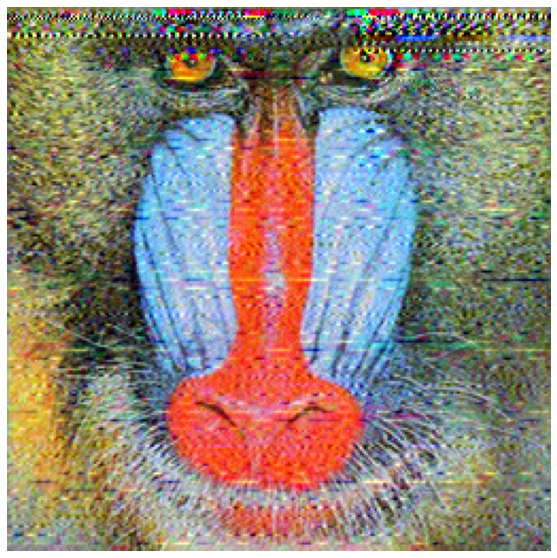}
    \includegraphics[width=.19\columnwidth]{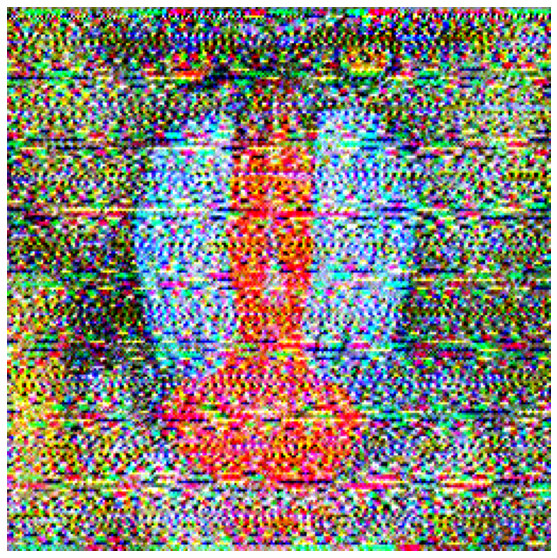}
    \includegraphics[width=.19\columnwidth]{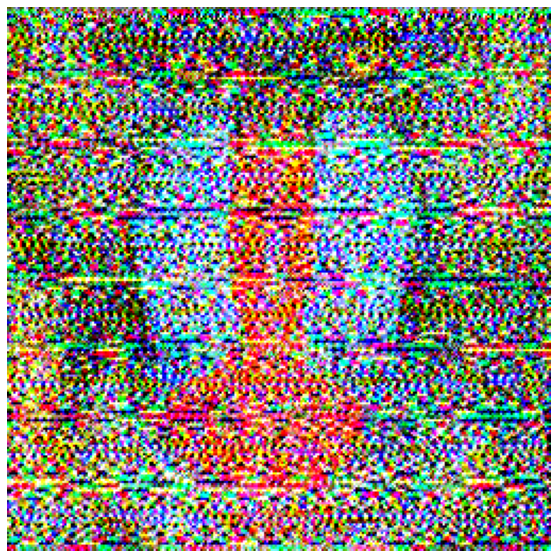}
    \includegraphics[width=.19\columnwidth]{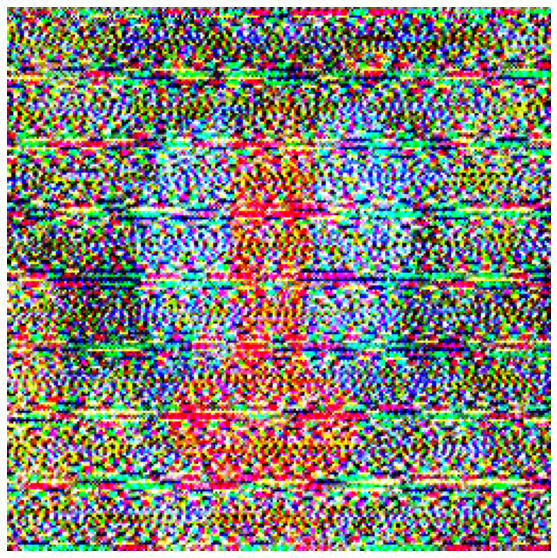}
    \caption[short]{Train: -, Test: AWGN.}
\end{subfigure}\\
\begin{subfigure}{\columnwidth}
    \centering
    \includegraphics[width=.19\columnwidth]{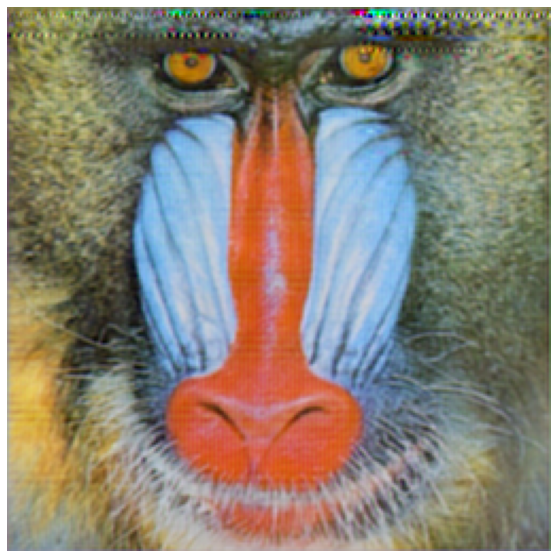}
    \includegraphics[width=.19\columnwidth]{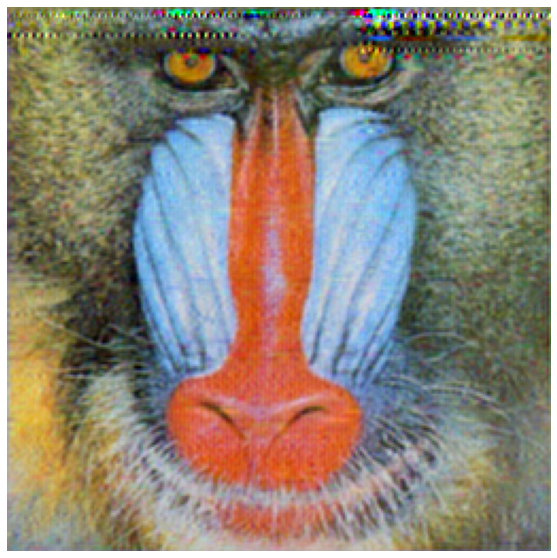}
    \includegraphics[width=.19\columnwidth]{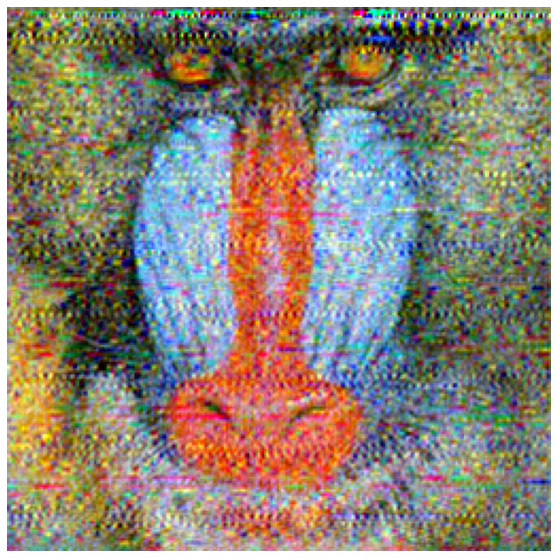}
    \includegraphics[width=.19\columnwidth]{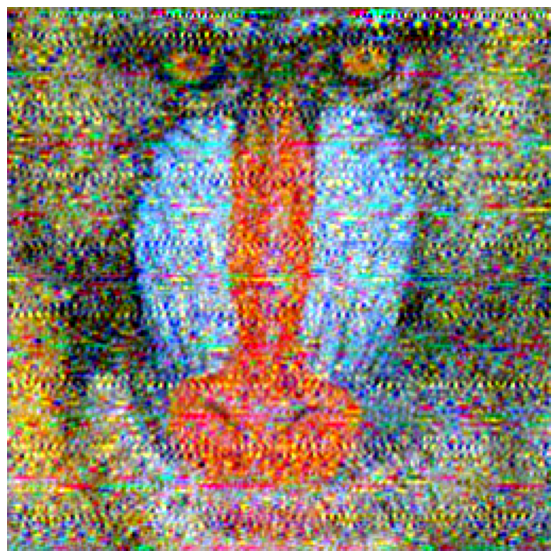}
    \includegraphics[width=.19\columnwidth]{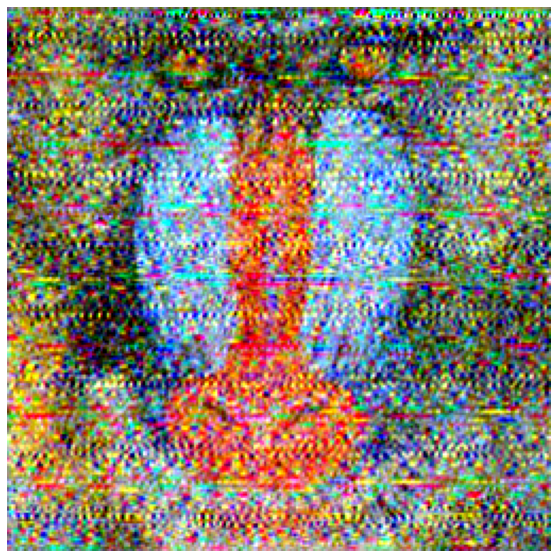}
    \caption[short]{Train: AWGN, Test: AWGN.}
\end{subfigure}\\
\begin{subfigure}{\columnwidth}
    \centering
    \includegraphics[width=.19\columnwidth]{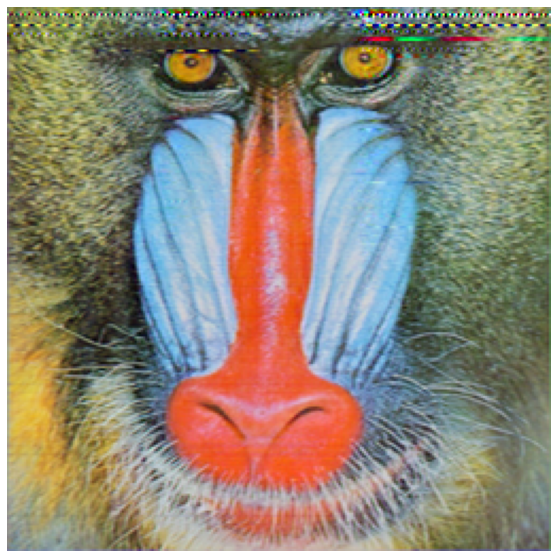}
    \includegraphics[width=.19\columnwidth]{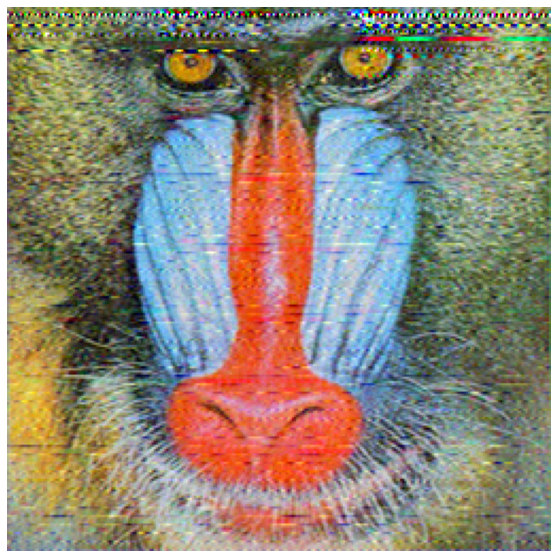}
    \includegraphics[width=.19\columnwidth]{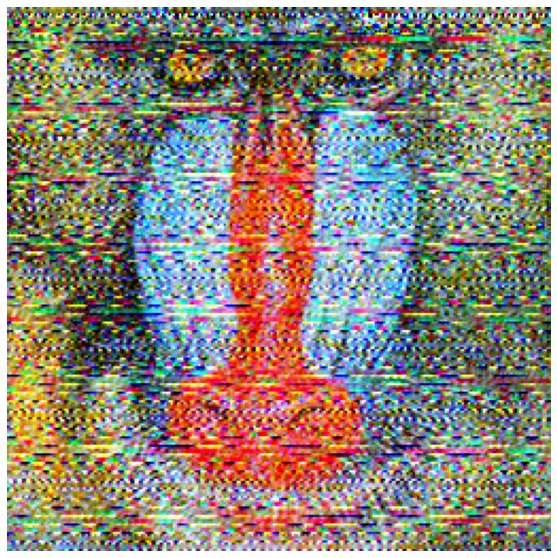}
    \includegraphics[width=.19\columnwidth]{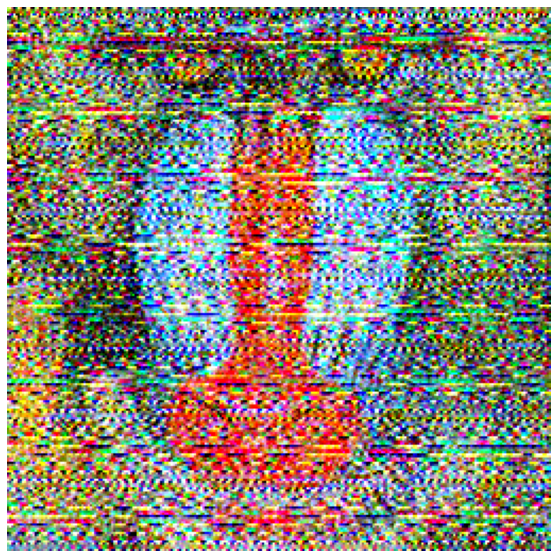}
    \includegraphics[width=.19\columnwidth]{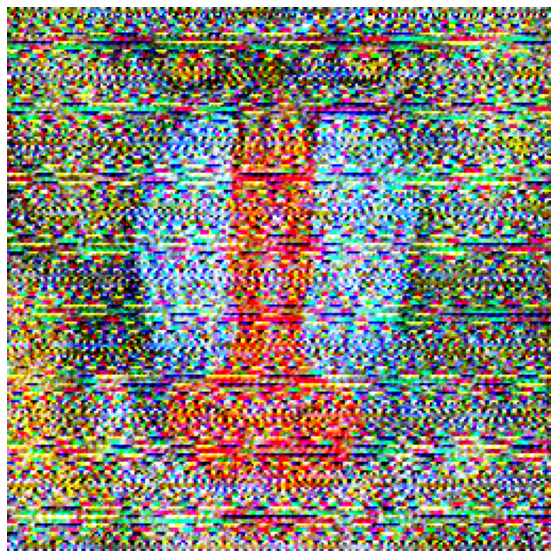}
    \caption[short]{Train: Speckle, Test: AWGN.}
\end{subfigure}\\
\begin{subfigure}{\columnwidth}
    \centering
    \includegraphics[width=.19\columnwidth]{figures/noise/2_tr_no_noise_vd_no_noise.png}
    \includegraphics[width=.19\columnwidth]{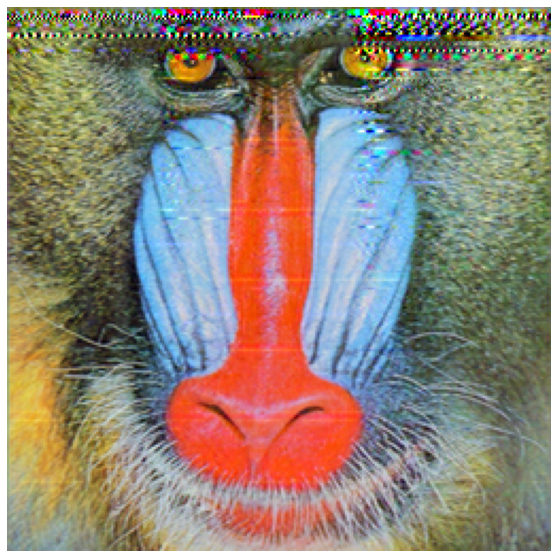}
    \includegraphics[width=.19\columnwidth]{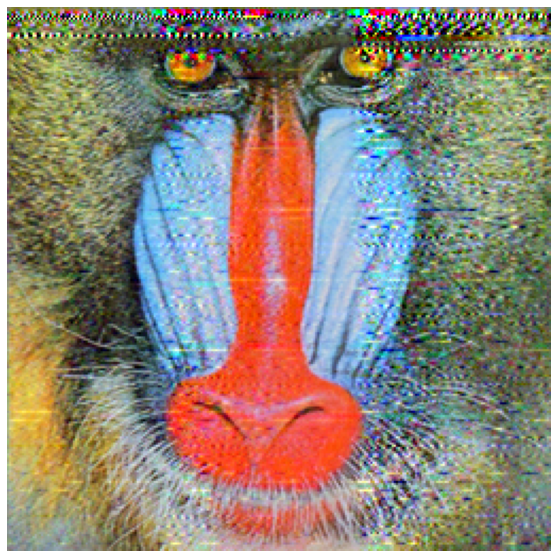}
    \includegraphics[width=.19\columnwidth]{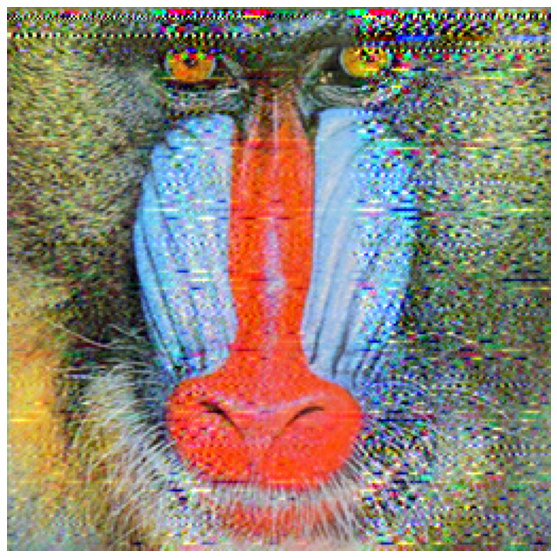}
    \includegraphics[width=.19\columnwidth]{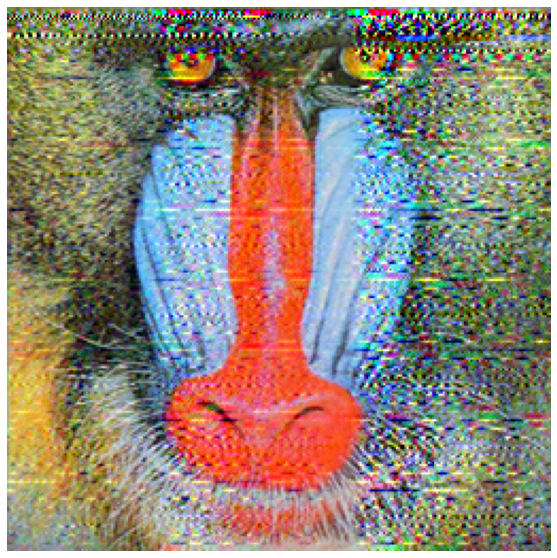}
    \caption[short]{Train: -, Test: Speckle.}
\end{subfigure}\\
\begin{subfigure}{\columnwidth}
    \centering
    \includegraphics[width=.19\columnwidth]{figures/noise/2_tr_gauss_0.1_vd_no_noise.png}
    \includegraphics[width=.19\columnwidth]{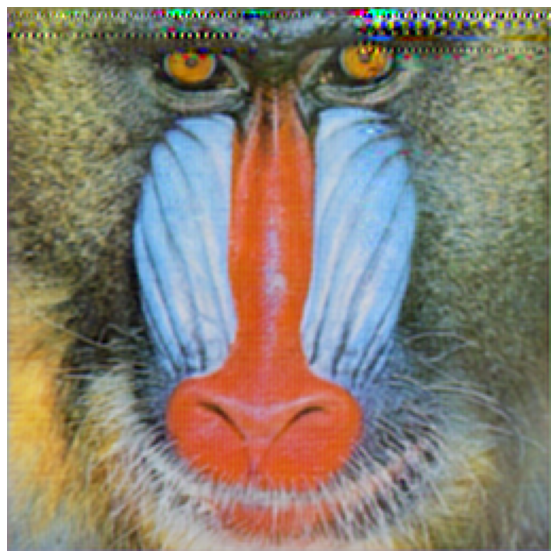}
    \includegraphics[width=.19\columnwidth]{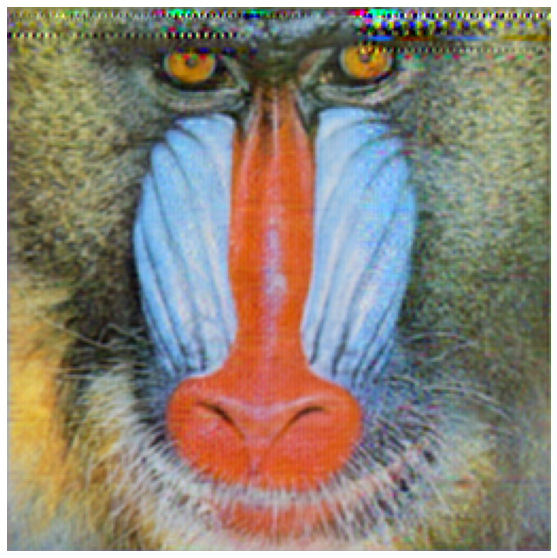}
    \includegraphics[width=.19\columnwidth]{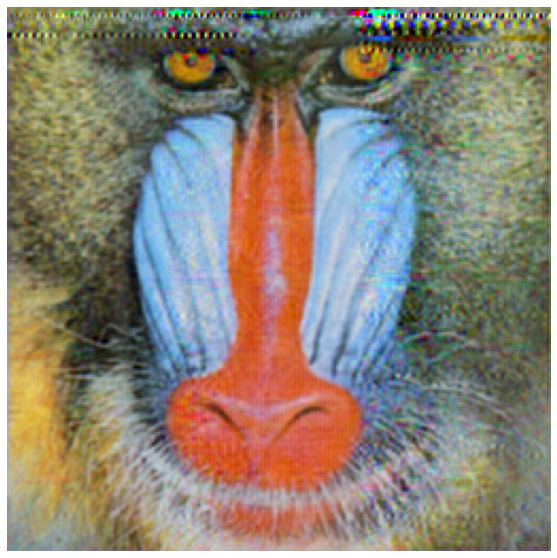}
    \includegraphics[width=.19\columnwidth]{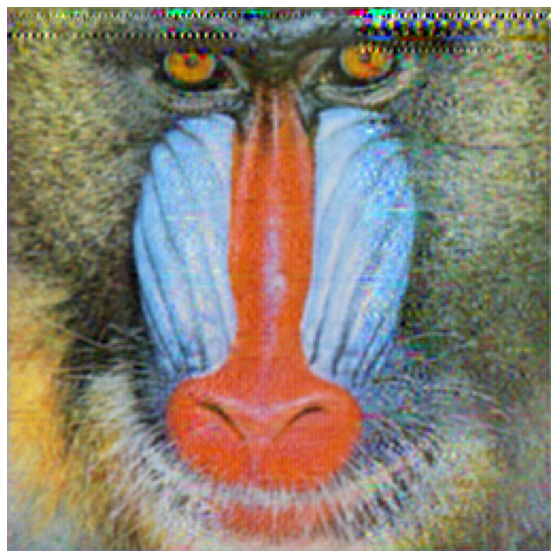}
    \caption[short]{Train: AWGN, Test: Speckle.}
\end{subfigure}\\
\begin{subfigure}{\columnwidth}
    \centering
    \includegraphics[width=.19\columnwidth]{figures/noise/2_tr_speckle_0.1_vd_no_noise.png}
    \includegraphics[width=.19\columnwidth]{figures/noise/2_tr_speckle_0.1_vd_gauss_0.1.png}
    \includegraphics[width=.19\columnwidth]{figures/noise/2_tr_speckle_0.1_vd_gauss_0.5.png}
    \includegraphics[width=.19\columnwidth]{figures/noise/2_tr_speckle_0.1_vd_gauss_0.75.png}
    \includegraphics[width=.19\columnwidth]{figures/noise/2_tr_speckle_0.1_vd_gauss_0.9.png}
    \caption[short]{Train: Speckle, Test: Speckle.}
\end{subfigure}\\
\caption[short]{\textbf{Robustness to AWGN and speckle noise:} Each column from left to right corresponds to an increasing $\sigma=$\{$0$, $0.1$, $0.5$, $0.75$, $0.9$\} The model trained with AWGN, depicted at rows (b) and (e), is the most robust, also under speckle noise that was never seen during training. The worst performance corresponds to the model with any noise at training time, depicted at rows (a) and (d).}

\label{fig:noise-qualitative}
\end{figure}

In a second stage, we study the impact of including synthetic AWGN or speckle noise during training time, to explore where they could bring a regularization effect that may increase the robustness of PixInWav.
The quantitative (Figure \ref{fig:noise-val}) and the qualitative (Figure \ref{fig:noise-qualitative}) results 
indicate that introducing AWGN at train time provides robustness against speckle noise.
This observation is somehow unexpected, because the model was expected to perform better when tested with the same type of noise with which it was trained.


\paragraph{Embedding capacity.} 
We are transmitting a 256 x 256 color image (3 channels) of 8 bits per pixel.
Each audio clip contains 67,522 samples at a sampling rate of 44,100 Hz, which corresponds to 1.53 seconds per clip. 
These values result in a transmission rate of 988 Kbps.

\paragraph{Computational cost of the method.} 
Both the hidden and reveal networks are identical and contain 482,090 parameters. The total computation cost of encoding, adding and decoding an image is of 197.31 GMAC (Giga multiply-accumulate operations).

\paragraph{Over the air transmission}

We developed a real deployment of \textit{PixInWav} by reproducing the raw waveform files from a loudspeaker, and recording this same sound from a microphone in a lab controlled setting.
The results presented in Table \ref{tab:onair} are promising, in the sense that some contours of the image are distinguishable, validating the feasibility of the proposed approach.
However, the quality of the image is very poor, so further research is needed to build noise resilient solutions.
To the best of our knowledge, this is the first time that pixels have been transmitted over the air with sound waves.

\begin{table}
\begin{center}
\resizebox{1.0\linewidth}{!}{
\begin{tabular}{ccc}
 Original & Recovered    & Misalignment\\
\includegraphics[width=.29\linewidth]{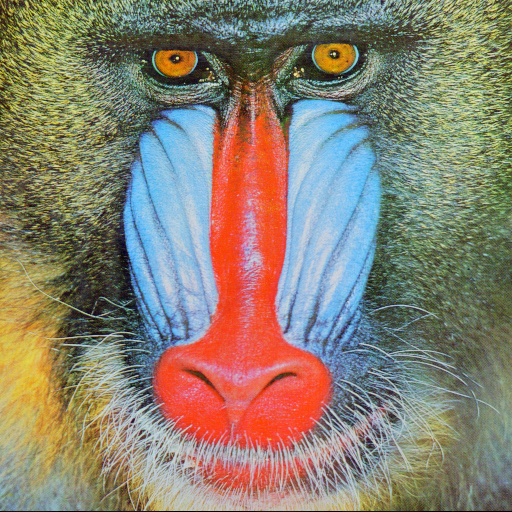} &
\includegraphics[width=.29\linewidth]{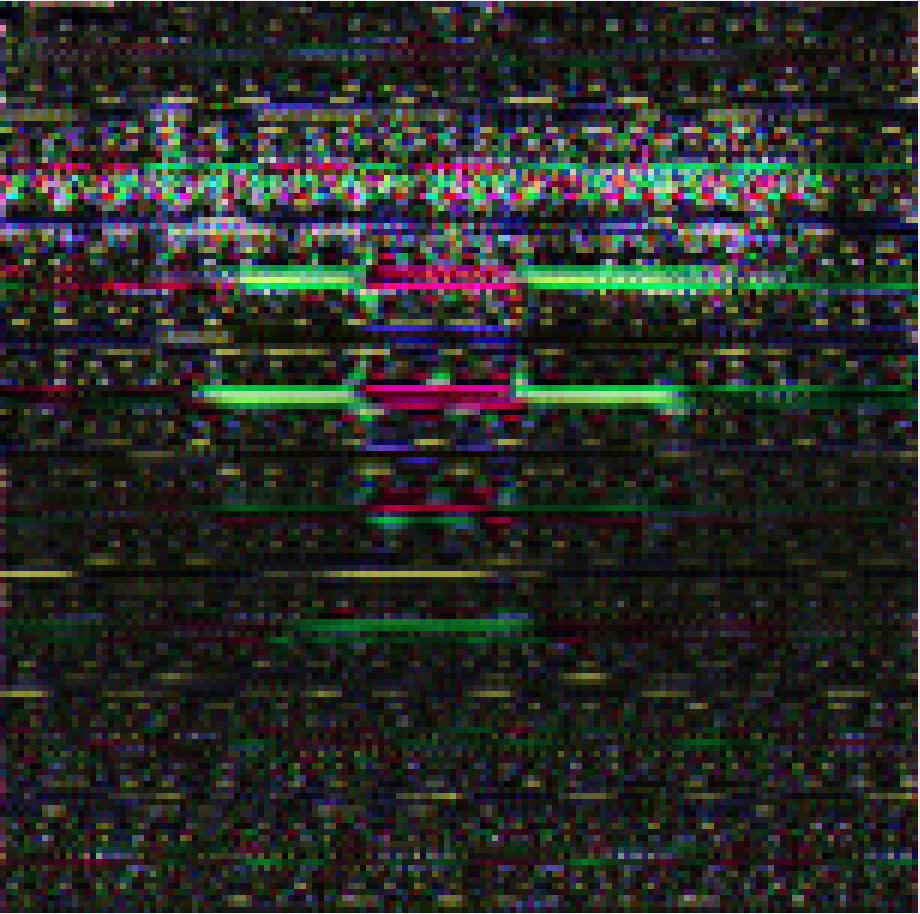}   &      \includegraphics[width=.29\columnwidth]{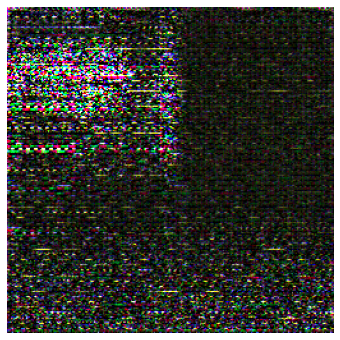} \\ 
\includegraphics[width=.29\linewidth]{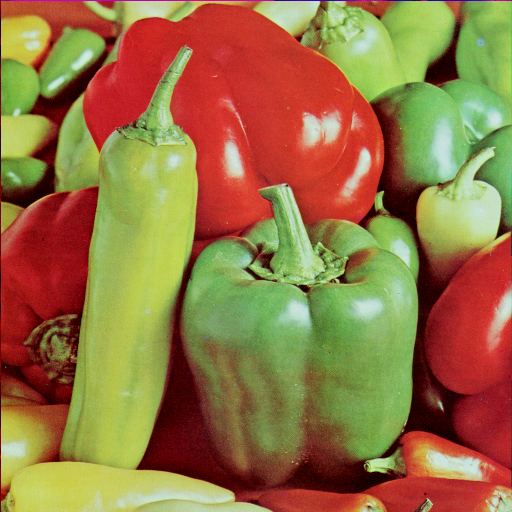} &
\includegraphics[width=.29\linewidth]{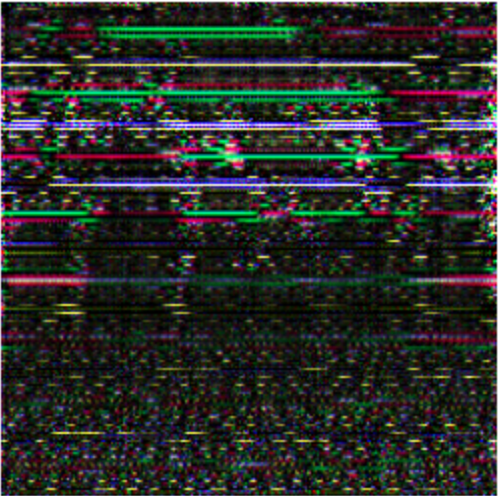}   &      \includegraphics[width=.29\columnwidth]{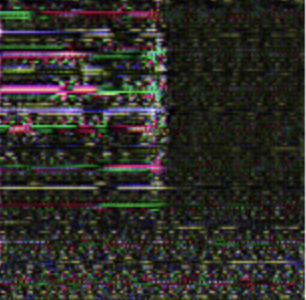} 
\end{tabular}}
\end{center}
\caption{\textbf{Image transmission over the air with sound waves:} The original image was encoded with \textit{PixInWav}, transmitted through the air by a loudspeaker, and recorded with a microphone. The recovered image (center) is of poor quality, but some pattern and even colors are distinguishable. In this experiment, the container signal was manually segmented and aligned to match the $1.5$ seconds of waveform. The other recovered image (right) shows a case in which this alignment is wrong, and where the vertical boundary of the image is clearly distinguishable from the portion of captured audio signal that does not transport any visual information.}
\label{tab:onair}
\end{table}

\section{Conclusions}
This paper presents pioneering work on deep multimodal steganography, in which we have explored the transmission of visual information over audio waveforms.
Our residual approach to deep steganography aims at obtaining an encoding of the hidden image that can be directly added to the host audio. Importantly, those hidden image encodings can be computed independently from the host audio --- what makes the system much more scalable.
We also found that the following strategies were beneficial: (i) training with the DTW audio loss, and (ii) employing the pixel shuffle layer for encoding the hidden image.
Finally, the residual and independent variation was been studied in detail for both Gaussian and speckle noises, finding that injecting Gaussian noise at training time actually improved the robustness against speckle noise.
This observation suggest that more elaborate training scheme considering the potential distortions of the audio during transmission could increase the robustness of the proposed PixInWav.




{\small
\bibliographystyle{ieee_fullname}
\bibliography{egbib}

\begin{thebibliography}{10}\itemsep=-1pt

\bibitem{atoum2013exploring}
Mohammed~Salem Atoum, Subariah Ibrahimn, Ghazali Sulong, Akram Zeki, and Adamu
  Abubakar.
\newblock Exploring the challenges of mp3 audio steganography.
\newblock In {\em 2013 International Conference on Advanced Computer Science
  Applications and Technologies}, pages 156--161. IEEE, 2013.

\bibitem{baluja2019hiding}
Shumeet Baluja.
\newblock Hiding images within images.
\newblock {\em IEEE transactions on pattern analysis and machine intelligence},
  42(7):1685--1697, 2019.

\bibitem{pmlr-v70-cuturi17a}
Marco Cuturi and Mathieu Blondel.
\newblock Soft-{DTW}: a differentiable loss function for time-series.
\newblock In Doina Precup and Yee~Whye Teh, editors, {\em Proceedings of the
  34th International Conference on Machine Learning}, volume~70 of {\em
  Proceedings of Machine Learning Research}, pages 894--903, International
  Convention Centre, Sydney, Australia, 06--11 Aug 2017. PMLR.

\bibitem{Cvejic}
N. Cvejic.
\newblock Algorithms for audio watermarking and steganography, 2004.
\newblock Department of Electrical and Information Engineering, Information
  Processing Laboratory, University of Oulu.

\bibitem{duan2019reversible}
Xintao Duan, Kai Jia, Baoxia Li, Daidou Guo, En Zhang, and Chuan Qin.
\newblock Reversible image steganography scheme based on a {U-Net} structure.
\newblock {\em IEEE Access}, 7:9314--9323, 2019.

\bibitem{ICSIspeechFAQ}
Dan Ellis.
\newblock Berkeley international computer science institute (icsi) speech faq,
  2009.

\bibitem{fonseca2019learning}
Eduardo Fonseca, Manoj Plakal, Daniel~PW Ellis, Frederic Font, Xavier Favory,
  and Xavier Serra.
\newblock Learning sound event classifiers from web audio with noisy labels.
\newblock In {\em ICASSP 2019-2019 IEEE International Conference on Acoustics,
  Speech and Signal Processing (ICASSP)}, pages 21--25. IEEE, 2019.

\bibitem{UED}
Linjie Guo, Jiangqun Ni, and Yun~Qing Shi.
\newblock An efficient jpeg steganographic scheme using uniform embedding.
\newblock In {\em 2012 IEEE International Workshop on Information Forensics and
  Security (WIFS)}, pages 169--174. IEEE, 2012.

\bibitem{he2016deep}
Kaiming He, Xiangyu Zhang, Shaoqing Ren, and Jian Sun.
\newblock Deep residual learning for image recognition.
\newblock In {\em Proceedings of the IEEE conference on computer vision and
  pattern recognition}, pages 770--778, 2016.

\bibitem{hmood2012new}
Dalal~N Hmood, Khamael~A Khudhiar, and Mohammad~S Altaei.
\newblock A new steganographic method for embedded image in audio file.
\newblock {\em International Journal of Computer Science and Security (IJCSS)},
  6(2):135--141, 2012.

\bibitem{WOW}
Vojt{\v{e}}ch Holub and Jessica Fridrich.
\newblock Designing steganographic distortion using directional filters.
\newblock In {\em 2012 IEEE International workshop on information forensics and
  security (WIFS)}, pages 234--239. IEEE, 2012.

\bibitem{S-UNIWARD}
Vojt{\v{e}}ch Holub, Jessica Fridrich, and Tom{\'a}{\v{s}} Denemark.
\newblock Universal distortion function for steganography in an arbitrary
  domain.
\newblock {\em EURASIP Journal on Information Security}, 2014(1):1, 2014.

\bibitem{pix2pix2017}
Phillip Isola, Jun-Yan Zhu, Tinghui Zhou, and Alexei~A Efros.
\newblock Image-to-image translation with conditional adversarial networks.
\newblock {\em CVPR}, 2017.

\bibitem{kreuk2020hide}
Felix Kreuk, Yossi Adi, Bhiksha Raj, Rita Singh, and Joseph Keshet.
\newblock Hide and speak: Towards deep neural networks for speech
  steganography.
\newblock {\em Proc. Interspeech 2020}, pages 4656--4660, 2020.

\bibitem{li2020invisible}
Shaofeng Li, Minhui Xue, Benjamin Zhao, Haojin Zhu, and Xinpeng Zhang.
\newblock Invisible backdoor attacks on deep neural networks via steganography
  and regularization.
\newblock {\em IEEE Transactions on Dependable and Secure Computing}, 2020.

\bibitem{LSB}
Jarno Mielikainen.
\newblock {LSB} matching revisited.
\newblock {\em IEEE signal processing letters}, 13(5):285--287, 2006.

\bibitem{pevny2010using}
Tom{\'a}{\v{s}} Pevn{\`y}, Tom{\'a}{\v{s}} Filler, and Patrick Bas.
\newblock Using high-dimensional image models to perform highly undetectable
  steganography.
\newblock In {\em International Workshop on Information Hiding}, pages
  161--177. Springer, 2010.

\bibitem{ronneberger2015u}
Olaf Ronneberger, Philipp Fischer, and Thomas Brox.
\newblock {U-Net}: Convolutional networks for biomedical image segmentation.
\newblock In {\em International Conference on Medical image computing and
  computer-assisted intervention}, pages 234--241. Springer, 2015.

\bibitem{russakovsky2015imagenet}
Olga Russakovsky, Jia Deng, Hao Su, Jonathan Krause, Sanjeev Satheesh, Sean Ma,
  Zhiheng Huang, Andrej Karpathy, Aditya Khosla, Michael Bernstein, et~al.
\newblock {ImageNet} large scale visual recognition challenge.
\newblock {\em International journal of computer vision}, 115(3):211--252,
  2015.

\bibitem{Venkataramani}
J.~Casebeer S.~Venkataramani and P. Smaragdis.
\newblock End-to-end source separation with adaptive front-ends, 2017.

\bibitem{santosa2005audio}
Rully~Adrian Santosa and Paul Bao.
\newblock Audio-to-image wavelet transform based audio steganography.
\newblock In {\em 47th International Symposium ELMAR, 2005.}, pages 209--212.
  IEEE, 2005.

\bibitem{shi2016real}
Wenzhe Shi, Jose Caballero, Ferenc Husz{\'a}r, Johannes Totz, Andrew~P Aitken,
  Rob Bishop, Daniel Rueckert, and Zehan Wang.
\newblock Real-time single image and video super-resolution using an efficient
  sub-pixel convolutional neural network.
\newblock In {\em Proceedings of the IEEE conference on computer vision and
  pattern recognition}, pages 1874--1883, 2016.

\bibitem{tancik2020stegastamp}
Matthew Tancik, Ben Mildenhall, and Ren Ng.
\newblock Stegastamp: Invisible hyperlinks in physical photographs.
\newblock In {\em Proceedings of the IEEE/CVF Conference on Computer Vision and
  Pattern Recognition}, pages 2117--2126, 2020.

\bibitem{tekeli2017comparison}
Kadir Tekeli and Rifat Asliyan.
\newblock A comparison of echo hiding methods.
\newblock {\em The Eurasia Proceedings of Science Technology Engineering and
  Mathematics}, 1:397--403, 2017.

\bibitem{wang2004image}
Zhou Wang, Alan~C Bovik, Hamid~R Sheikh, and Eero~P Simoncelli.
\newblock Image quality assessment: from error visibility to structural
  similarity.
\newblock {\em IEEE transactions on image processing}, 13(4):600--612, 2004.

\bibitem{DBLP:journals/corr/abs-1907-06956}
Mehdi Yedroudj, Fr{\'{e}}d{\'{e}}ric Comby, and Marc Chaumont.
\newblock Steganography using a 3 player game.
\newblock {\em CoRR}, abs/1907.06956, 2019.

\bibitem{zhu2018hidden}
Jiren Zhu, Russell Kaplan, Justin Johnson, and Li Fei-Fei.
\newblock Hidden: Hiding data with deep networks.
\newblock In {\em Proceedings of the European conference on computer vision
  (ECCV)}, pages 657--672, 2018.

\end{thebibliography}
}

\end{document}